\documentclass[11pt]{article}
\usepackage{graphicx}
\usepackage{bm}
\usepackage{amsmath}
\usepackage{amssymb}
\usepackage{amsfonts}
\usepackage{amsthm}
\usepackage{mathtools}
\usepackage{hyperref}
\usepackage{braket}
\usepackage[normalem]{ulem}
\usepackage{wrapfig}
\usepackage{cite}
\usepackage{tikz}
\usepackage[left=1in,right=1in,top=1in,bottom=1in]{geometry}

\hypersetup{colorlinks=true,urlcolor=[rgb]{0,0,0.5},citecolor=[rgb]{0.5,0,0},linkcolor=[rgb]{0,0,0.4}}

\newtheorem*{definition}{Definition}

\newtheorem*{conjecture}{Conjecture}

\renewcommand{\title}[1]{\vbox{\center\bf{\Large #1}}\vspace{5mm}}
\renewcommand{\author}[1]{\vbox{\center{#1}}\vspace{5mm}}
\newcommand{\address}[1]{\vbox{\center\em#1}}

\newcommand\emails[1]{\begingroup
	\renewcommand\thefootnote{}\footnote{#1}
	\addtocounter{footnote}{-1}\endgroup}

\def\Tr{{\rm Tr}}

\def\op{{\cal O}}

\def\iden{\mathbb{I}}
\def\1den{\hbox{$1\hskip -1.2pt\vrule depth 0pt height 1.53ex width 0.7pt
                  \vrule depth 0pt height 0.3pt width 0.12em$}}
\def\with{\quad {\rm with} \quad}
\def\where{\quad {\rm where} \quad}

\def\and{\quad {\rm and} \quad}

\def\ni{\noindent}
\def\nn{\nonumber\\}

\def\ra{\rightarrow}
\def\ie{{\rm i.e.\ }}
\def\eg{{\rm e.g.\ }}

\def\CA{{\cal A}}

\def\CE{{\cal E}}
\def\CF{{\cal F}}
\def\CH{{\cal H}}

\def\Wg{{\cal W}\! g}
\def\ktitle{\ensuremath{k}}

\begin{document}

\begin{titlepage}

\begin{center}
\vspace*{2cm}
\title{Unitary designs from statistical mechanics\\  \vspace*{4pt} in random quantum circuits}
\author{Nicholas Hunter-Jones}
\address{Perimeter Institute for Theoretical Physics,\\ Waterloo, Ontario N2L 2Y5, Canada}

\emails{\hspace*{-8mm} \href{mailto:nickrhj@perimeterinstitute.ca}{\tt nickrhj@perimeterinstitute.ca} }

\end{center}

\begin{abstract}
Random quantum circuits are proficient information scramblers and efficient generators of randomness, rapidly approximating moments of the unitary group. We study the convergence of local random quantum circuits to unitary $k$-designs. Employing a statistical mechanical mapping, we give an exact expression of the distance to forming an approximate design as a lattice partition function. In the statistical mechanics model, the approach to randomness has a simple interpretation in terms of domain walls extending through the circuit. We analytically compute the second moment, showing that random circuits acting on $n$ qudits form approximate $2$-designs in $O(n)$ depth, as is known. 
Furthermore, we argue that random circuits form approximate unitary $k$-designs in $O(nk)$ depth and are thus essentially optimal in both $n$ and $k$. We can show this in the limit of large local dimension, but more generally rely on a conjecture about the dominance of certain domain wall configurations.
\end{abstract}

\end{titlepage}

\tableofcontents

\section{Introduction}

Random quantum circuits are invaluable constructions in both quantum information and quantum many-body physics. They are a rare example of a solvable model of strongly-coupled dynamics and are efficient implementations of randomness.
From a quantum information perspective, random quantum circuits are low-depth constructions of unitary $k$-designs \cite{HL08,BHH12}, rapid information scramblers \cite{HaydenPreskill,BF12}, and essentially optimal decouplers \cite{BF13}. Much has been understood about their convergence properties \cite{ELL05,ODP07,HL08,BHH12,Zni08,BV10,BH13}. Moreover, recent benchmarks for demonstrating quantum advantage involve the complexity of sampling the output distribution of random circuits \cite{Boixo18}. From a many-body viewpoint, quantum circuits are simple models of local chaotic dynamics and are a valuable resource in elucidating the onset of scrambling and thermalization in quantum systems. Recently, random circuits have been used to study both operator growth \cite{NVH17,vonKey17} and the spreading of entanglement \cite{NRVH16,NahumQuench17,RQCstatmech} under chaotic evolution. 

In this work we study the convergence of local random quantum circuits to unitary $k$-designs, ensembles which emulate fully random unitaries by capturing the first $k$ moments of the Haar measure on the unitary group, and thus yielding the power of random unitaries with much lower complexity. Designs are central in quantum information theory, with applications permeating a broad set of subfields. Prior work has shown that random circuits form approximate $k$-designs in a circuit depth that scales linearly in $n$ and polynomially in $k$ \cite{BHH12}. Our approach entails a synthesis of ideas from quantum information and condensed matter physics, utilizing an exact mapping to the statistical mechanics of a lattice model. These techniques for random circuits were studied in \cite{NVH17,RQCstatmech}, and we make extensive use of the ideas developed there.

We focus on the frame potential, measuring the 2-norm distance to Haar-randomness, and show that frame potential can be exactly written as the partition function of a spin system on a triangular lattice. For $k=2$, the local degrees of freedom in the lattice model are simple and the approach to forming a 2-design can be understood in terms of decaying domain wall configurations. The second frame potential is essentially exactly computable. We find that random circuits form $\epsilon$-approximate 2-designs in a depth $n+\log(1/\epsilon)$ and precisely determine the coefficients.

The statistical mapping is exact for general $k$, but the combinatorics of domain wall counting in the lattice model for higher moments are less amenable to precise calculation. There is a single type of non-intersecting domain wall in the second moment, but in higher moments we have complex configurations of different types of interacting domain walls. We prove some general properties of the $k$-th partition function, which allows us to compute the leading order domain wall contribution. In the limit of large local dimension, we show that the $k$-design depth is $O(nk)$. We can be precise in this limit, but for general local dimension a rigorous bound on the higher order terms that contribute to the partition function is not at hand. Nevertheless, given a conjecture that a simple sector of domain wall terms dominates the multidomain wall terms, which we provide evidence for, we argue that random quantum circuits form approximate $k$-designs in depth $nk+k\log k + \log(1/\epsilon)$. As the lower bound on the depth required to form a $k$-design is linear in $n$ and $k$, this implies that random circuits are essentially optimal implementations of unitary $k$-designs. 

We start by overviewing random quantum circuits and presenting definitions of the frame potential and the notion of an approximate $k$-design. In Sec.~\ref{sec:statmech}, we describe an exact mapping which allows us to write the frame potential for random circuits as the partition function of a triangular lattice model. We then give a precise treatment of the second moment in Sec.~\ref{sec:2design}, and understand the non-zero configurations that contribute to the partition function in terms of domain walls in the lattice model. In Sec.~\ref{sec:kdesigns}, we extend the discussion to general moments and argue that random circuits form $k$-designs in $O(nk)$ depth.

\subsubsection*{Random quantum circuits}
\begin{figure}
\centering
\includegraphics[width=0.45\linewidth]{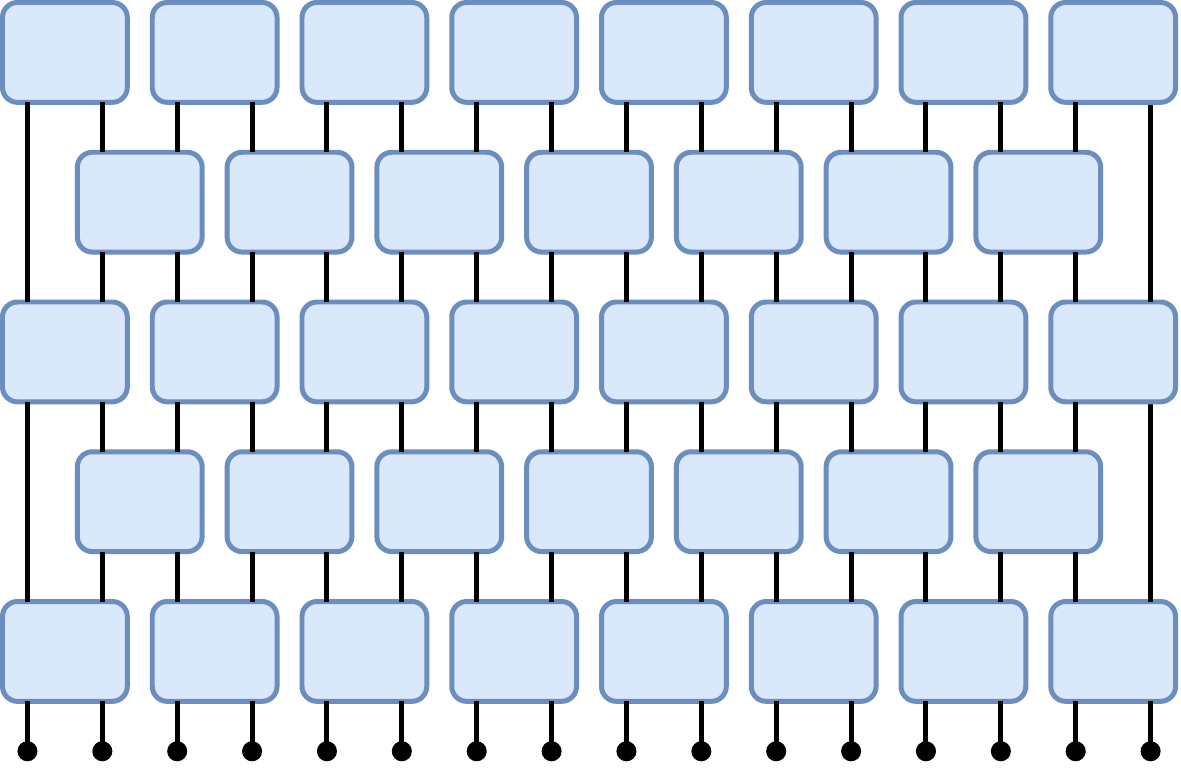}
\begin{tikzpicture}[thick,scale=0.6,baseline=-0.4cm]
\draw[thick,->] (0,0) -- (0,6) node[anchor=south] {$t$};
\end{tikzpicture}
\caption{The random circuits we consider are built from staggered layers of 2-site unitaries on $n$ qudits of local dimension $q$, where each gate is drawn randomly from $U(q^2)$.}
\label{fig:RQCdiag}
\end{figure}
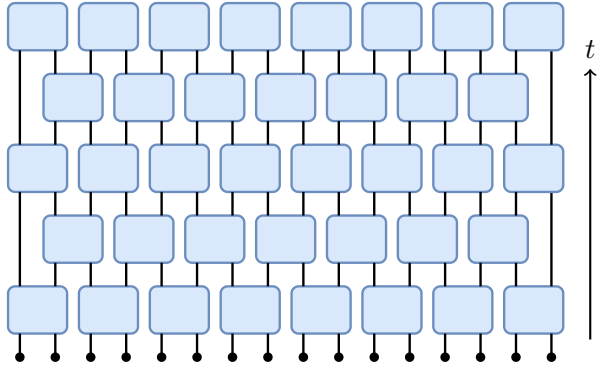

The random quantum circuits (RQCs) we consider act on 1-dimensional chains of $n$ qudits with local dimension $q$. We evolve in time by acting with staggered layers of 2-site random unitaries, acting on even links at even time steps and odd links at odd time steps, and where each 2-site unitary is drawn Haar randomly from $U(q^2)$. Explicitly, even time steps are given by the tensor product of 2-site unitaries $\bigotimes_i U_{i,i+1}$ for even $i$, and odd time steps by $\bigotimes_i U_{i,i+1}$ for odd $i$. Time evolution to time $t$ is given by acting with $t$ layers of unitaries, such that the geometry of the circuit is fixed and not random, as shown in Fig.~\ref{fig:RQCdiag}.

Our goal is to compute the frame potential for random circuits, where the $k$-th frame potential quantifies the 2-norm distance between between the $k$-fold channels. Focusing on the 2-norm will allow for exact calculation. The approach will be to give analytic expressions for the 2-norm and then bound the approximate design depth from the diamond distance between the $k$-fold channels. We will restrict ourselves to considering moments less than the dimension $k\leq d=q^n$ and note that this is sufficient as $O(d^2)$ gates is enough to implement a fully Haar random unitary.

We first give a brief summary of the quantum information theoretic tools we need, including the definition of an approximate $k$-design and the frame potential. A more detailed discussion with the necessary definitions and proofs of some of the statements below is given in App.~\ref{app:designs}. 

\subsubsection*{Approximate $k$-designs}
The Haar measure is the unique left/right invariant measure on the unitary group $U(d)$. We are interested in when a set of unitaries captures moments of the full unitary group. Consider an ensemble of unitaries $\CE$, a subset of $U(d)$ equipped with a probability measure. For an operator $\op$ acting on the $k$-fold Hilbert space $\CH^{\otimes k}$, the $k$-fold channel with respect to $\CE$ is defined as
\begin{equation}
\Phi^{(k)}_{\CE} (\op) = \int_\CE dU\, U^{\otimes k}(\op)U^\dagger{}^{\otimes k}\,.
\end{equation}
In this sense we ask when the average of an operator over the ensemble $\CE$ equals an average over the full unitary group. A {\it unitary $k$-design} is an ensemble $\CE$ for which the $k$-fold channels are equal for all operators $\op$,
\begin{equation}
\Phi^{(k)}_{\CE} (\op) = \Phi^{(k)}_{\rm Haar} (\op)\,.
\end{equation}
Forming a $k$-design means we exactly capture the first $k$ moments of the Haar measure, and reproducing higher and higher moments corresponds to better approximating the full unitary group. For $k=1$, any basis for the algebra of operators of $\CH$, including the Pauli group, is a 1-design. But for general $k$, very few constructions of exact $k$-designs are known, with the exception of $k=2$ and $k=3$ \cite{Dankert09,Cleve16,Kueng15,Webb15,Zhu15}. Instead, one may relax this notion and ask when an ensemble of unitaries is {\it close} to forming a $k$-design. 

\begin{definition}[approximate $k$-design] 
\label{def:kdesign}
For $\epsilon>0$, an ensemble of unitaries $\CE$ is an $\epsilon$-approximate $k$-design if the diamond norm of the difference of $k$-fold channels is
\begin{equation}
\big\|\Phi^{(k)}_\CE - \Phi^{(k)}_{\rm Haar}\big\|_\diamond \leq \epsilon\,.
\end{equation}
\end{definition}
There are varying definitions of approximate designs, some of which involve different norms, but the definitions are equivalent up to powers of the dimension; see \cite{LowThesis} for a nice overview. Lastly, we note that \cite{BHH12} used a slightly stronger definition 
in terms of the complete positivity of the difference in superoperators, but this will only affect our bounds by an additional factor of the dimension.

\subsubsection*{Frame potential}
The frame potential for an ensemble of unitaries $\CE$ was first discussed in the context of unitary designs in \cite{Gross07,Scott08}. More recent interest involves the frame potential as a diagnostic of chaotic dynamics \cite{ChaosDesign,ChaosRMT,ChaosSUSYSYK,Zhuang19}. Specifically, \cite{ChaosDesign} related the frame potential to both averaged out-of-time-ordered correlators and measures of complexity, and \cite{ChaosRMT} computed the frame potentials for random Hamiltonian evolution.

\begin{definition}[frame potential]
The $k$-th frame potential of an ensemble of unitaries $\CE$ is defined as a double average over the ensemble
\begin{equation}
\CF^{(k)}_\CE = \int_{U,V\in \CE} dUdV\,\big| \Tr(U^\dagger V)|^{2k}\,.
\label{eq:FPdef}
\end{equation}
\end{definition}
\ni The frame potential for any ensemble is lower bounded by its Haar value as
\begin{equation}
\CF^{(k)}_\CE \geq \CF^{(k)}_{\rm Haar}\,,
\label{eq:FPbound}
\end{equation}
with equality if and only if the ensemble forms a $k$-design. The Haar value of the frame potential is $\CF^{(k)}_{\rm Haar} = k!$ for moments $k\leq d$.

Relating this measure of Haar-randomness to our definition of an approximate design, the difference in the $k$-th frame potentials of an ensemble $\CE$ and the Haar ensemble bounds the diamond norm of the difference in $k$-fold channels as
\begin{equation}
\big\|\Phi^{(k)}_\CE - \Phi^{(k)}_{\rm Haar}\big\|_\diamond^2 \leq d^{2k} \big(\CF^{(k)}_\CE-\CF^{(k)}_{\rm Haar}\big)\,.
\label{eq:FPdiam}
\end{equation}
We present the proofs of both of the above statements in Eq.~\eqref{eq:FPbound} and Eq.~\eqref{eq:FPdiam} in App.~\ref{app:designs}. 

\subsubsection*{Previous results}
Harrow and Low \cite{HL08} (with \cite{DinizComm11}) studied local random quantum circuits with randomly applied gates, and by looking at the mixing time of the Markov chain on Pauli strings, showed that they form approximate $2$-designs in $O(n^2)$ depth. In their random circuits each time step corresponds to the application of a single gate, whereas in our parallelized random circuits each time step consists of $O(n)$ gates, and thus the time scales differ by a factor of $n$. 

These results were extended in \cite{BH13} by studying the gap of the moment operator, showing that random circuits form 3-designs. The gap of the second moment operator was also computed exactly in \cite{Zni08}, and properties of the $k$-th moment operator were studied in \cite{BV10}, both for 2-local all-to-all coupled circuits. 

Brand{\~a}o, Harrow, and Horodecki \cite{BHH12,BHH16} further extended this approach to general $k$, showing that random circuits form approximate unitary $k$-designs in a depth $O(n k^{10})$. More recently, \cite{HM18} studied higher dimensional random circuits, demonstrating that they form $k$-designs in depth $O(n^{1/D} {\rm poly}(k))$, with the degree of the polynomial depending on the dimension $D$ of the lattice. Furthermore, it was shown in \cite{BHH12} that the lower bound on the depth required for a 1D random circuit to form a $k$-design is linear in $n$ and $k$, so the dependence on $k$ cannot be improved by more than polynomial factors. Progress towards optimality in $k$ was made in \cite{Nakata16} for random Hamiltonian evolution, where they found convergence to $k$-designs in $O(n^2 k)$ steps, for moments up to $k = o(\sqrt{n})$. It has also been shown \cite{Onorati17} that Brownian quantum circuits \cite{FastScrambling} form designs in $O(n k^{10})$ depth, with the same dependence on $k$ as in \cite{BHH12}. Lastly, we note that \cite{kdesignnum13} numerically investigated the gap of the moment operator and found evidence for a linear growth in design from the first few moments.

\subsubsection*{Our results}
We show that the frame potential for random quantum circuits on $n$ qudits of local dimension $q$ and circuit depth $t$ can be exactly written as a partition function of a triangular lattice model 
\vspace*{-4pt}
\begin{equation}
\CF^{(k)}_{\rm RQC} = \sum_{\{\sigma\}} \begin{gathered}\includegraphics[width=2.5cm]{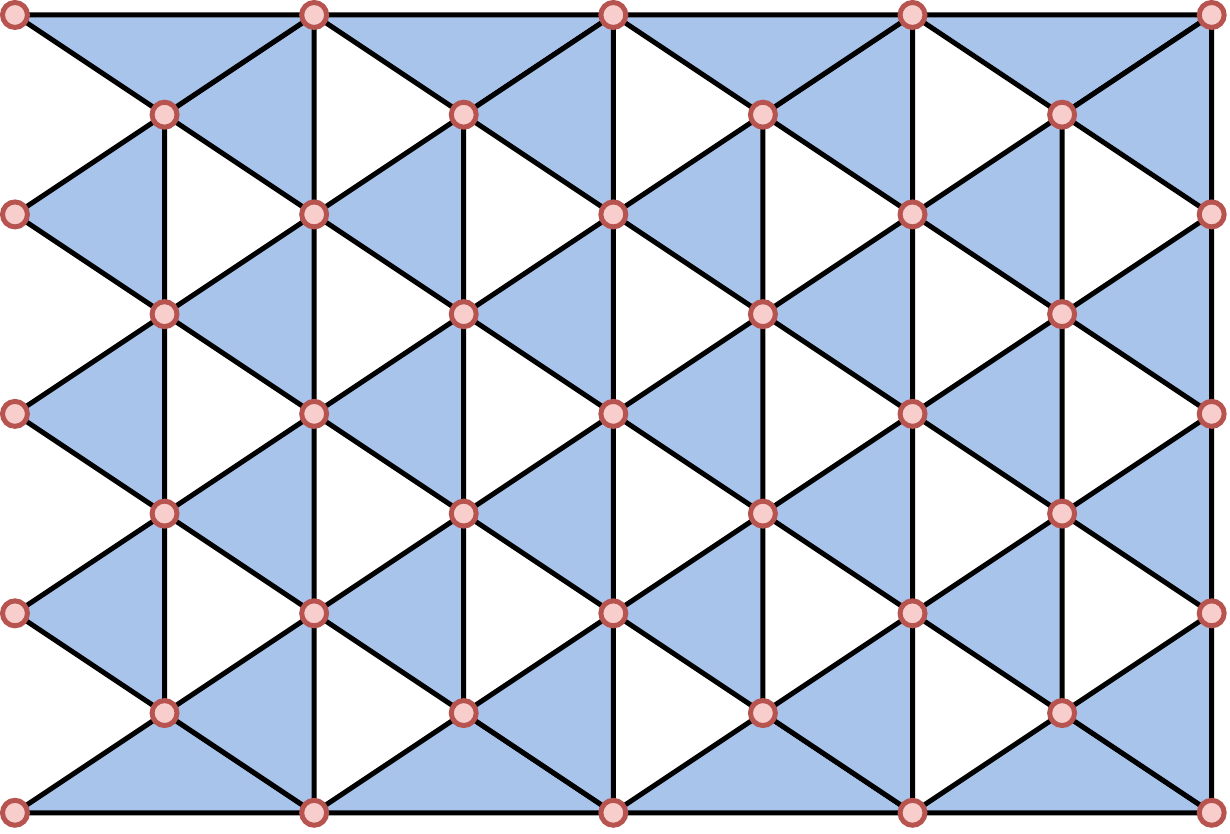}\end{gathered} \vspace*{-6pt}
\end{equation}
where the local degrees of freedom are permutations $\sigma \in S_k$, and the lattice is of width $n_g = \lfloor n/2\rfloor $, depth $2(t-1)$, and has periodic boundary conditions in time. We emphasize that the lattice model should simply be viewed as a tool to analytically compute the frame potential. 

For the second moment, the $k=2$ frame potential is essentially exactly computable and the nonzero spin configurations have a simple interpretation in terms of domain walls separating regions of local identity and swap permutations. By counting the domain wall configurations we compute the depth at which random circuits form $\epsilon$-approximate unitary $2$-designs to be
\begin{equation}
t_2 \geq C\big( 2n \log q + \log n + \log 1/\epsilon\big) \with C = \bigg(\log\frac{q^2+1}{2q}\bigg)^{-1}\,,
\end{equation}
with the linear term $\approx 6.2 n$ for $q=2$ and asymptoting to $2n$ for $q\ra \infty$. 

The partition function for general $k$ receives contributions from more complicated domain wall configurations, where each domain wall represents an element of the generating set of transpositions for the symmetric group $S_k$. We prove some general properties for the plaquette terms constituting the $k$-th partition function and show that only a simple sector of domain wall configurations contributes at leading order in $1/q$, which allows us to compute the depth at which we form a $k$-design in the large $q$ limit to be 
\begin{equation}
t_k = O(nk)\,.
\end{equation}
Furthermore, we argue that, given a conjecture that the sector of single domain wall configurations bounds the multi-domain wall configurations at any finite $q$, random quantum circuits are optimal generators of randomness achieving the lower bound on the design depth.

\section{Exact lattice mappings for RQCs}
\label{sec:statmech}
Random unitary circuits admit an exact mapping to a classical spin system, allowing for a simplification in computation and analytic treatment of the moments. A classical mapping for random tensor networks was first discussed in \cite{RTN16}, but explicit details for the circuits we consider here were described in \cite{NVH17,RQCstatmech}. Specifically, \cite{NVH17} used the statistical mechanical mapping to exactly compute the out-of-time ordered correlation function in random circuits, and \cite{RQCstatmech} extended the techniques for a replica calculation of the R\'enyi entropies. 

The quantity of interest here is the $k$-th frame potential for random unitary circuits, defined as
\begin{equation}
\CF^{(k)}_{\rm RQC} = \int_{U,V\in {\rm RQC}} dUdV\, \big|\Tr(U_t^\dagger V_t)|^{2k}\,,
\end{equation}
for RQCs evolved to time $t$. Visualizing this as a circuit diagram, for a single $U^\dagger V$ we have
\begin{equation}
\begin{aligned}
\includegraphics[width=8cm]{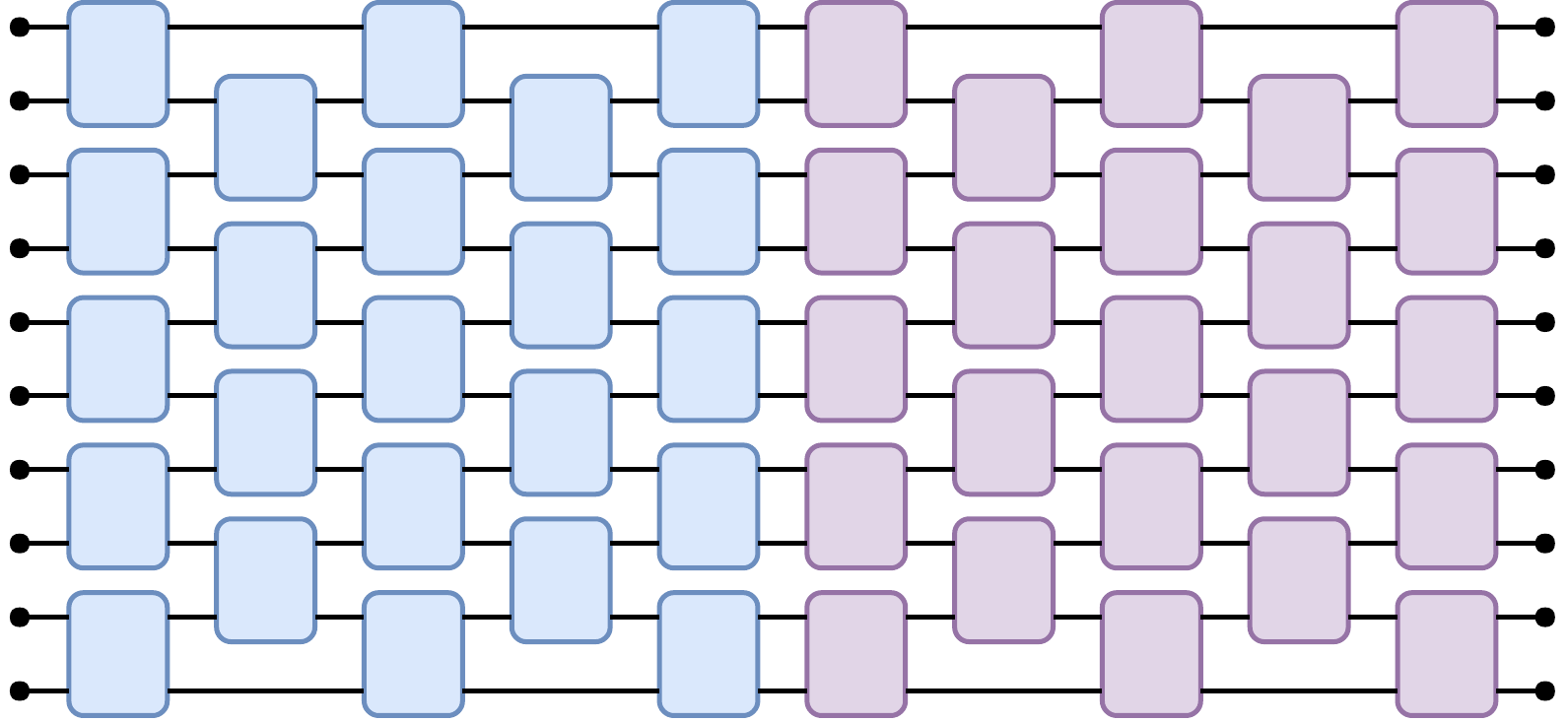}
\end{aligned}
\label{eq:UVdiag}
\end{equation}
with the $U$ time increasing left to right on the blue circuit and the $V$ time increasing right to left on the purple circuit. The trace enforces periodic boundary conditions in time in the circuit. For the $k$-th frame potential we have $k$ copies of $U$ and $V$ and their complex conjugates. First note that we can combine a single layer on each side of the $U$ and $V$ circuits using the left/right-invariance of the Haar measure, \ie in the middle and through the trace. As we are averaging over each 2-site unitary in the $U$ and $V$ circuits independently, the frame potential is equivalently an average of moments of traces of depth $2(t-1)$ circuits
\begin{equation}
\CF^{(k)}_{\rm RQC} = \int_{\rm RQC} dU\, \big|\Tr(U_{2(t-1)})|^{2k}\,.
\end{equation}

We want to consider $k$-th moments of the random circuit and compute averages over all 2-site unitaries. If we imagine stacking the $k$ copies of the circuit and its conjugate on top of each other, then we see that performing an average over each 2-site unitary of the form $U^{\otimes k} \otimes U^\dagger{}^{\otimes k}$ will give index contractions with neighboring gates on the $k$-fold space.

We review Haar integration in more detail in App.~\ref{app:designs}, but briefly recall that averages over polynomials of random unitaries take the form
\begin{equation}
 \int dU\, U_{\vec \imath, \vec \jmath }^{\otimes k} \otimes U^\dagger{}^{\otimes k}_{\vec \ell, \vec m} = \sum_{\sigma, \tau \in S_k} \delta_\sigma(\vec \imath\,| \vec m)\delta_\tau(\vec \jmath\,| \vec \ell\,) \Wg(\sigma^{-1} \tau, d)\,,
 \end{equation} 
where the $i$'s and $j$'s are the indices of the $U$'s, and $\ell$'s and $m$'s are the indices of the $U^\dagger$'s. Heuristically, Haar integration is a prescription for index contraction, telling us to contract in the ingoing indices of $U$ to the outgoing indices of $U^\dagger$ and the outgoing indices of $U$ to the ingoing indices of $U^\dagger$, both indexed by a permutation with a weight that is a function of the permutations. Explicitly $\delta_\sigma (\vec \imath\, |\vec \jmath\,) = \delta_{i_1,j_{\sigma(1)}}\ldots \delta_{i_k,j_{\sigma(k)}}$ and $\Wg(\sigma,d)$ is the Weingarten function on elements of the symmetric group $S_k$. 

Thinking about each gate stacked on top of the $2k$ copies of itself, and denoting the gates as $U_{ij}$ and $U^\dagger_{\ell m}$, the Weingarten formula then tells us to contract $i \ra m$ and $j\ra \ell$ indices, each with respect to an element of $S_k$. We can represent this with an effective vertex \cite{NVH17}:
\begin{center}
\begin{tikzpicture}[thick,scale=0.55]
\node at (0,0) {\includegraphics[width=5cm]{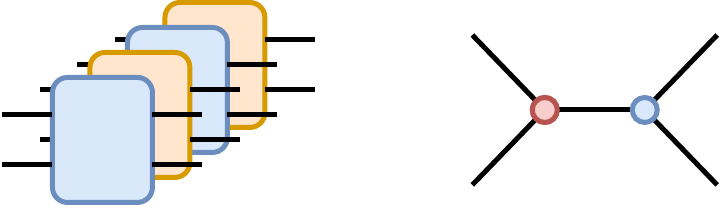}};
\draw[thick,->] (0,0) -- (0.8,0);
\end{tikzpicture}
\end{center}
The blue gates are $U$'s and the yellow gates are $U^*$'s (\ie from folding the circuit of $U^\dagger$'s). We have incoming $i m$ indices and outgoing $j\ell$ indices, and by averaging the gate we generate index contractions in each of the legs of the vertex with respect to a permutation at the node. 

The red nodes are permutations $\sigma \in S_k$, the blue nodes are $\tau \in S_k$, the horizontal lines are weights given by the Weingarten function $\Wg(\sigma^{-1}\tau, q^2)$, and the diagonal lines denote the index contractions between the legs of two gates in the circuit. These contractions between legs give a weight determined by an inner product between permutations: $\braket{\sigma|\tau} = q^{\ell(\sigma^{-1}\tau)}$, where $\ell(\sigma^{-1}\tau)$ is the length of the cycle-type of the permutation product, \ie the number of closed loops in the product. For $k=2$ we have
\begin{equation}
\braket{\sigma|\tau} = 
\begin{tikzpicture}[thick,scale=0.6,baseline=0.5cm]
\draw[rounded corners] (0,0.1) rectangle (0.4,1.9);
\draw[rounded corners] (1.8,0.1) rectangle (2.2,1.9);
\foreach \y in {0.4,0.8,1.2,1.6}
{\draw (0.4,\y) -- (1.8,\y);}
\node at (0.2,-0.15) {$\sigma$};
\node at (2,-0.15) {$\tau$};
\end{tikzpicture}
= q^{\ell(\sigma^{-1}\tau)}\,,
\end{equation}
where the four lines are the contractions between gates in the two copies of the circuit and its conjugate, and each permutation gives contraction between the two top lines and the two bottom lines, giving factors of the local dimension $q$. For instance, the inner products we find for $k=2$ are: $\braket{\iden|\iden} = q^2$, $ \braket{S|S} = q^2 $, $\braket{\iden|S} $, and $ \braket{S|\iden} = q$.

\begin{figure}
\centering
\includegraphics[width=6.8cm]{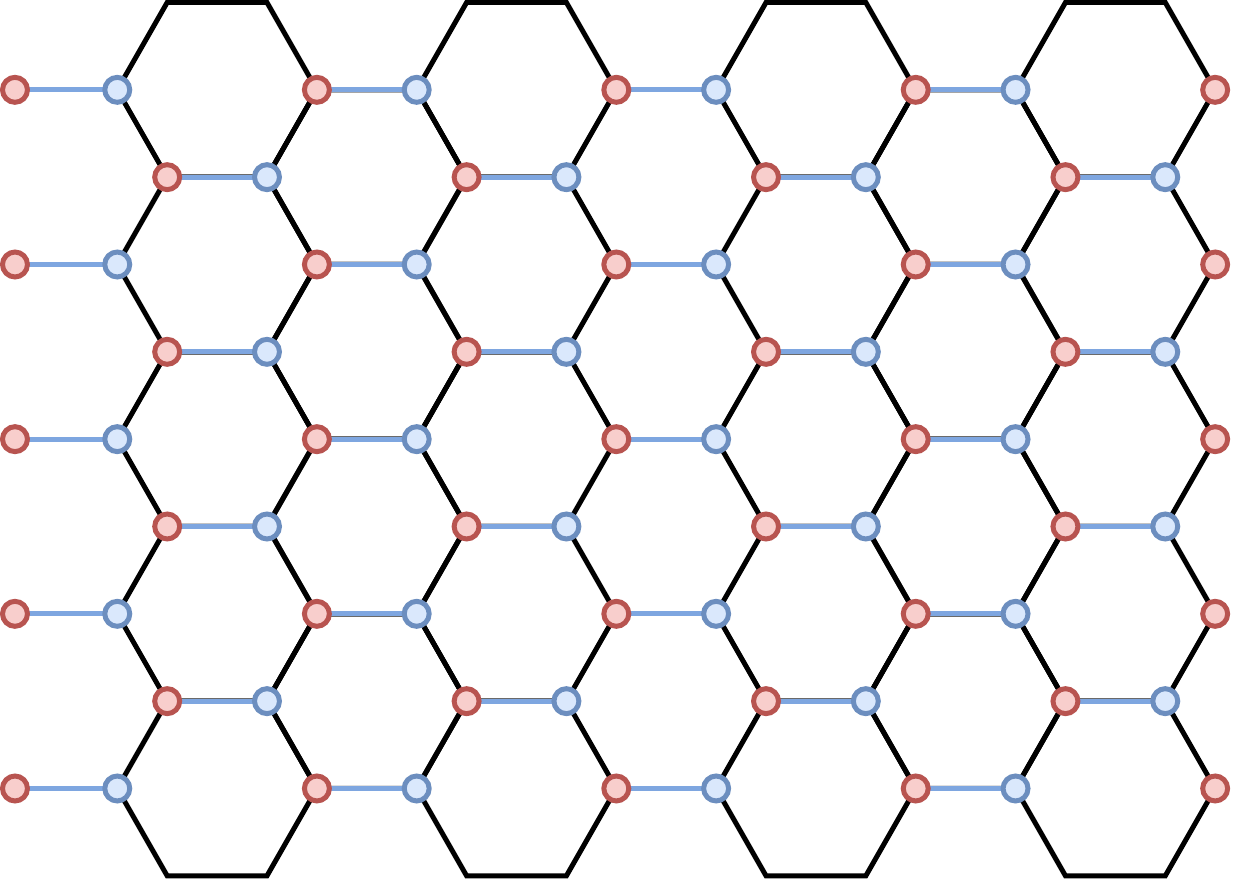}
\qquad \includegraphics[width=6.8cm]{RQCtri}
\caption{The $k$-th frame potential can be written as the partition function of a spin system on a hexagonal lattice with local $S_k$ spins (left). By summing over the blue nodes we can define effective plaquette terms and write the frame potential as the partition function on a triangular lattice (right). In both figures time runs from left to right and periodic boundary conditions in time means the red nodes on the ends of the circuit are identified.}
\label{fig:FPlattice}
\end{figure}

Integrating over each 2-site unitary in the circuit means we replace each gate with the effective vertex above and sum over all $\sigma, \tau \in S_k$. The result is that the $k$-th frame potential can be written as the partition function of a hexagonal lattice model on the left in Fig.~\ref{fig:FPlattice}, summing over permutations in $S_k$ at each node and assigning a weight as described above. We have a depth $2(t-1)$ lattice with periodic boundary conditions in time, so the leftmost and rightmost red nodes in the lattice are identified. 

The $k$-th frame potential is exactly equal to a sum over spin configurations on a hexagonal lattice.\footnote{We will refer to the local degrees of freedom in the lattice model as spins, keeping in mind they are elements of the symmetric group $S_k$.}
A priori, this reduction to an spin system is not too enlightening. We must evaluate all spin configurations, many of which are negative, and carefully assign weights to a given configuration as prescribed above. But, as was described in \cite{NVH17}, some substantial simplifications occur when we sum over certain local spins.

Performing the sum over the blue nodes in the partition function, the local $\tau$ spins, we can then define effective plaquette terms, which are functions of three permutations
\begin{equation}
\begin{tikzpicture}[thick,scale=0.5,baseline=-0.1cm]
\node[anchor=east,inner sep=0] at (-2,0) {\includegraphics[width=4.5cm]{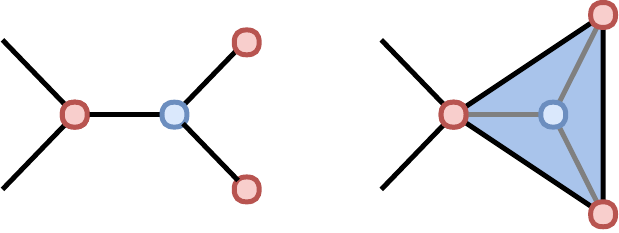}};
\draw[thick,->] (-6.8,0) -- (-6,0);
\end{tikzpicture} \quad\where 
J^{\sigma_1}_{\sigma_2 \sigma_3} \equiv 
\sum_{\tau\in S_k}
\begin{tikzpicture}[thick,scale=0.48,baseline=-0.1cm]
\node at (0,0) {\includegraphics[width=1.3cm]{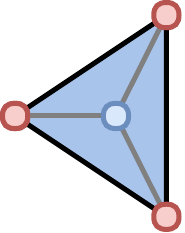}};
\node[anchor=east] at (-1.2,0) {$\sigma_1$};
\node[anchor=west] at (1.2,1.5) {$\sigma_2$};
\node[anchor=west] at (1.2,-1.5) {$\sigma_3$};
\node[anchor=west] at (0.3,0.05) {$\tau$};
\end{tikzpicture}
\end{equation}
thereby reducing frame potential to a partition function on a triangular lattice as shown in Fig.~\ref{fig:FPlattice}.\footnote{Had we not used the invariance of $U(q^2)$ to absorb a layer of gates between $U$ and $V^\dagger$ in Eq.~\eqref{eq:UVdiag}, we would have had a boundary vertex between the two circuits, where summing over the $\tau$ spins on the blue node gives a $\delta$-function for $\sigma$ spins
\vspace*{-8pt}
\begin{equation*}
\begin{tikzpicture}[baseline=-0.08cm]\node at (0,0) {\includegraphics[width=1.4cm]{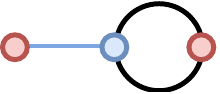}};\end{tikzpicture}
\quad \ra \quad
\begin{tikzpicture}[baseline=-0.08cm]\node at (0,0) {\includegraphics[width=1cm]{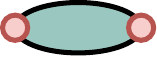}};\end{tikzpicture} = \delta_{\sigma_1 \sigma_2}\,.
\end{equation*}
\vspace*{-12pt}
}

With this reduction, the frame potential is now equal to the partition function of local $S_k$ spins on a triangular lattice
\begin{equation}
\CF^{(k)}_{\rm RQC} = \sum_{\{\sigma\}} \prod_{\triangleleft} J^{\sigma_1}_{\sigma_2 \sigma_3} = \sum_{\{\sigma\}} \begin{gathered}\includegraphics[width=3cm]{RQCtri}\end{gathered}
\end{equation}
where the lattice is $n_g = \lfloor n/2 \rfloor$ sites across, with $n_g$ denoting the number of gates in a layer, and $2(t-1)$ sites deep, with periodic boundary conditions in time. The plaquettes are functions of three permutations, written explicitly in terms of the Weingarten functions and permutation inner products as
\begin{equation}
J^{\sigma_1}_{\sigma_2\sigma_3} =
\begin{tikzpicture}[thick,scale=0.55,baseline=-0.05cm]
\node at (0,0) {\includegraphics[width=1cm]{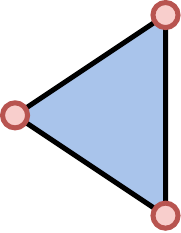}};
\node[anchor=east] at (-0.8,0) {$\sigma_1$};
\node[anchor=west] at (0.8,1.1) {$\sigma_2$};
\node[anchor=west] at (0.8,-1.1) {$\sigma_3$};
\end{tikzpicture}
=\sum_{\tau \in S_k} \Wg(\sigma_1^{-1}\tau,d) q^{\ell(\tau^{-1}\sigma_2)} q^{\ell(\tau^{-1}\sigma_3)}\,.
\label{eq:Jplaq}
\end{equation}

We stress that the lattice partition function is not physical, but is simply a way to analytically express moments of the random circuit. Furthermore, we note that the random circuit quantities one wishes to compute are entirely encoded in the boundary conditions of the lattice model, whether they be correlation functions \cite{NVH17} or R\'enyi entropies \cite{RQCstatmech}. In this sense, the frame potential is simple as it merely enforces periodic boundary conditions in the lattice model. 

\subsubsection*{$k=2$ plaquette terms}
We reduced the computation of the frame potential to a combinatorial problem of enumerating lattice configurations of spins. In the $k=2$ partition function the local spins are $\sigma\in S_2 = \{\iden, S\}$, the identity and swap permutations. Using Eq.~\eqref{eq:Jplaq}, we can compute the possible plaquette terms for the partition function of the second moment \cite{NVH17}:
\begin{equation}
\begin{aligned}
\begin{tikzpicture}[thick,scale=0.55,baseline=-0.05cm]
\node at (0,0) {\includegraphics[width=1cm]{Jvertex}};
\node[anchor=east] at (-0.8,0) {$\iden$};
\node[anchor=west] at (0.8,1.1) {$\iden$};
\node[anchor=west] at (0.8,-1.1) {$\iden$};
\end{tikzpicture} = 1\,, 
\qquad
\begin{tikzpicture}[thick,scale=0.55,baseline=-0.05cm]
\node at (0,0) {\includegraphics[width=1cm]{Jvertex}};
\node[anchor=east] at (-0.8,0) {$\iden$};
\node[anchor=west] at (0.8,1.1) {$S$};
\node[anchor=west] at (0.8,-1.1) {$S$};
\end{tikzpicture} = 0\,,&
\qquad
\begin{tikzpicture}[thick,scale=0.55,baseline=-0.05cm]
\node at (0,0) {\includegraphics[width=1cm]{Jvertex}};
\node[anchor=east] at (-0.8,0) {$\iden$};
\node[anchor=west] at (0.8,1.1) {$\iden$};
\node[anchor=west] at (0.8,-1.1) {$S$};
\end{tikzpicture} = 
\begin{tikzpicture}[thick,scale=0.55,baseline=-0.05cm]
\node at (0,0) {\includegraphics[width=1cm]{Jvertex}};
\node[anchor=east] at (-0.8,0) {$\iden$};
\node[anchor=west] at (0.8,1.1) {$S$};
\node[anchor=west] at (0.8,-1.1) {$\iden$};
\end{tikzpicture} =\frac{q}{(q^2+1)}\,,\\
\begin{tikzpicture}[thick,scale=0.55,baseline=-0.05cm]
\node at (0,0) {\includegraphics[width=1cm]{Jvertex}};
\node[anchor=east] at (-0.8,0) {$S$};
\node[anchor=west] at (0.8,1.1) {$S$};
\node[anchor=west] at (0.8,-1.1) {$S$};
\end{tikzpicture} = 1\,,
\qquad
\begin{tikzpicture}[thick,scale=0.55,baseline=-0.05cm]
\node at (0,0) {\includegraphics[width=1cm]{Jvertex}};
\node[anchor=east] at (-0.8,0) {$S$};
\node[anchor=west] at (0.8,1.1) {$\iden$};
\node[anchor=west] at (0.8,-1.1) {$\iden$};
\end{tikzpicture} = 0\,,&
\qquad
\begin{tikzpicture}[thick,scale=0.55,baseline=-0.05cm]
\node at (0,0) {\includegraphics[width=1cm]{Jvertex}};
\node[anchor=east] at (-0.8,0) {$S$};
\node[anchor=west] at (0.8,1.1) {$S$};
\node[anchor=west] at (0.8,-1.1) {$\iden$};
\end{tikzpicture} = 
\begin{tikzpicture}[thick,scale=0.55,baseline=-0.05cm]
\node at (0,0) {\includegraphics[width=1cm]{Jvertex}};
\node[anchor=east] at (-0.8,0) {$S$};
\node[anchor=west] at (0.8,1.1) {$\iden$};
\node[anchor=west] at (0.8,-1.1) {$S$};
\end{tikzpicture}= \frac{q}{(q^2+1)}\,.&
\end{aligned}
\label{eq:k2rules}
\end{equation}

Plaquette terms in the first column mean that lattice configurations of all $\iden$'s or all $S$'s have weight one. Moreover, the plaquette terms in the second column ensure that any spin configuration with a single permutation differing from the rest will have weight zero. We can understand the possible contributions to the partition function by thinking about domain walls separating regions of identity and swap permutations. 
The nonzero plaquette terms in Eq.~\eqref{eq:k2rules} mean that, propagating through the network in the time direction, a domain wall can either move left or right at a plaquette. Given the periodic boundary conditions in time, the non-zero subleading contributions in the partition function arise from domain walls between $\iden$'s and $S$'s running through the circuit
\begin{equation}
\CF^{(2)}_{\rm RQC} = 2 + \sum \mbox{domain wall configurations}\,,
\end{equation}
where each domain wall contributes a factor of $q/(q^2+1)$ for each time step. We see that not only must we sum over single domain walls, but also configurations of multiple domain walls, as shown in Fig.~\ref{fig:RQCdw}. Moreover, Eq.~\eqref{eq:k2rules} guarantees that domain walls cannot end, and thus, given the periodic boundary conditions, no closed loops are allowed. The same reasoning holds for the boundary plaquettes; domain walls may enter through the boundary, but cannot exit. Periodic boundary conditions ensure that all such configurations are zero.

\subsubsection*{\ktitle-designs from the ground states}
In this picture it is clear where $k!$, the minimal Haar-random value of the frame potential, comes from. The plaquette terms have weight one when evaluated on the same spin $\sigma \in S_k$. We discuss this in Sec.~\ref{sec:kdesigns} and prove this for general $k$ in App.~\ref{app:kplaq}. Therefore, in the $k$-th frame potential there are $k!$ configurations with unit weight. In the language of the spin partition function, there is an energy cost associated to having different neighboring spins. The ground states of the lattice model are the $k!$ spin configurations with all sites are equal. The excited states correspond to domains of differing spins, where there is an energy cost in $1/q$ for differing neighboring spins. We will give a more extensive discussion of the $k$-th frame potential in Sec.~\ref{sec:kdesigns}, where we compute the contributions from excited states,
but in the strict limit of large local dimension $q\ra \infty$, we see that $\CF^{(k)}_{\rm RQC} \approx k! +O(1/q)$.

\subsubsection*{Some exact results for RQCs}
We briefly comment on some straightforward exact results. The frame potential for RQCs at time $t=0$ is simply the contribution from the identity $\CF^{(k)}_{\rm RQC}(t=0) = d^{2k} = q^{2nk}$. At time $t=1$, with one layer of circuit evolution, we have a tensor product of 2-site unitaries. The frame potential is a product of the moments of traces of the gates $\CF^{(k)}_{\rm RQC}(t=1) = (k!)^{n/2}$, for moments $k\leq q^2$ and assuming an even number of sites. For $k>q^2$, the moments of traces of the 2-site unitaries are known combinatorial factors.

The first moment of random circuits is also simple. Every gate we Haar average generates a single index contraction with its neighboring gates, factors of the dimension exactly canceled by the Weingarten coefficients. In the language of our lattice model, there is only one local degree of freedom in $S_1$ and only one kind of plaquette term with weight 1. The $k=1$ frame potential is
\begin{equation}
\CF^{(1)}_{\rm RQC}(t) = 1\,,
\end{equation}
and thus random quantum circuits form {\it exact 1-designs} for all $q$, $n$, and $t$. 

\begin{figure}
\centering
\includegraphics[width=4.5cm]{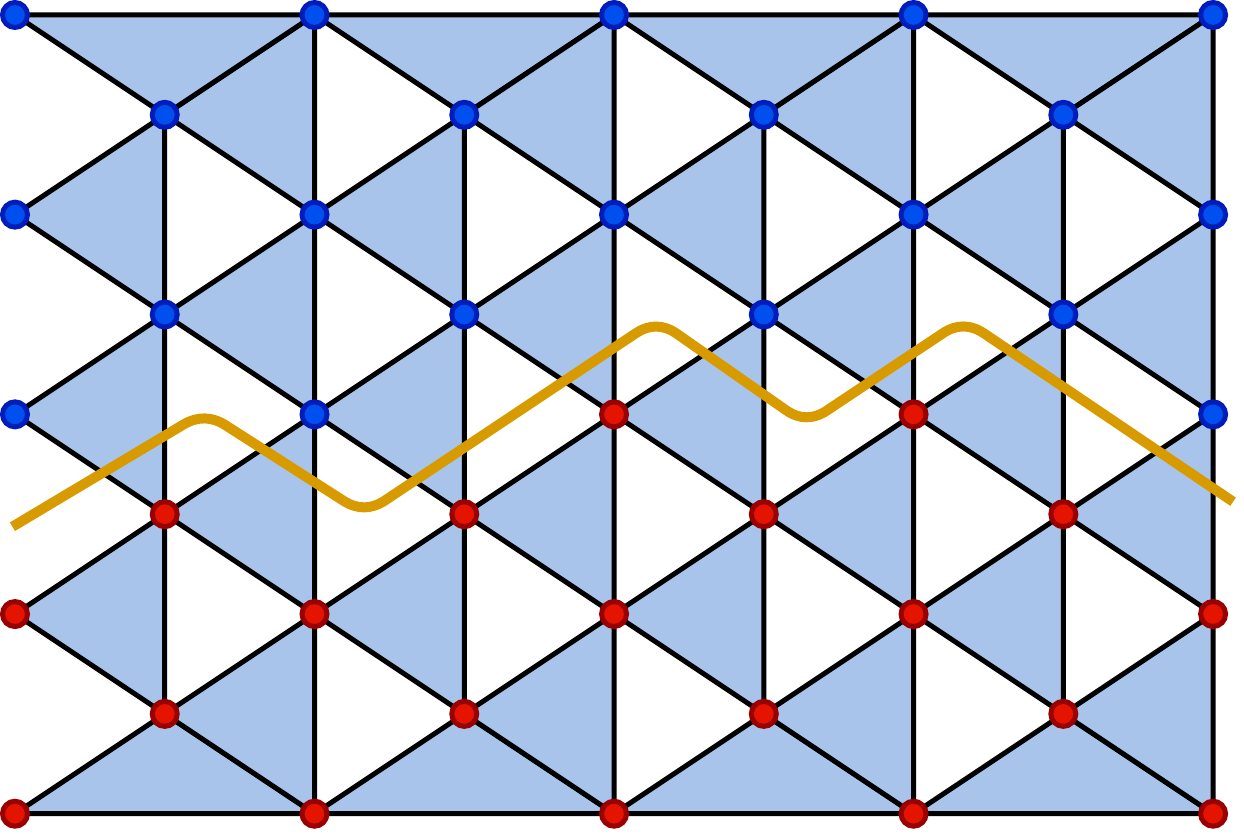} \qquad\qquad \includegraphics[width=4.5cm]{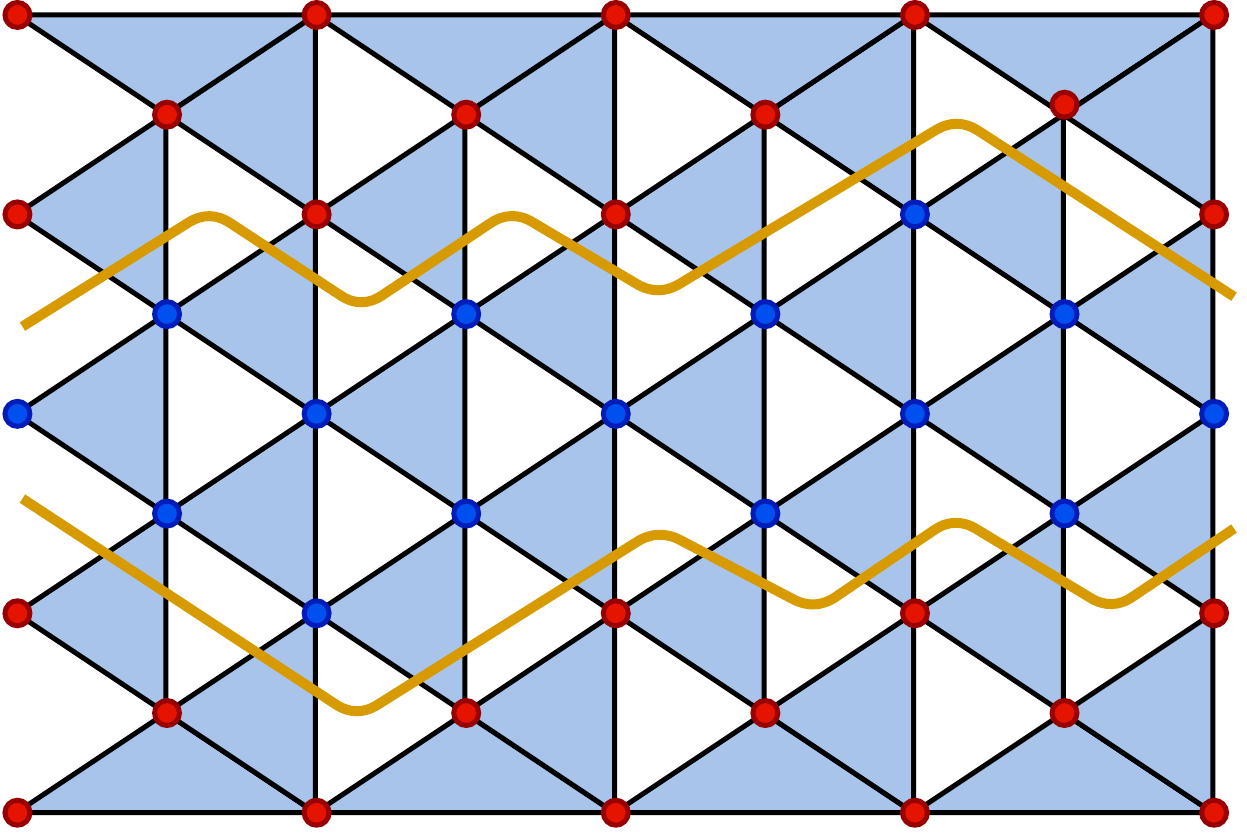}
\caption{Examples of nonzero contributions to the $k=2$ partition function: a single and double domain wall configuration, where the domain walls separate regions of local identity and swap permutations.}
\label{fig:RQCdw}
\end{figure}

\section{2-designs from statistical mechanics}
\label{sec:2design}
Computing the $k$-th frame potential exactly from the statistical mechanics model might be difficult, but for $k=2$, where we have local Ising-like spins and simple domain walls, an analytic treatment is more straightforward. As we showed, the $k=2$ frame potential for random quantum circuits is simply the partition function for $S_2$ spins on a directed triangular plaquette model. We computed the exact plaquette terms in Eq.~\eqref{eq:k2rules} and argued that the $k=2$ partition function is a sum over all domain wall configurations on the triangular lattice
\begin{equation}
\CF^{(2)}_{\rm RQC} = 2\bigg( 1 + \sum_{\rm 1~dw} wt(q,t) + \sum_{\rm 2~dw} wt(q,t) +\ldots\bigg)\,.
\end{equation}
Two such possible configurations are shown in Fig.~\ref{fig:RQCdw}. We can upper bound this quantity directly. There is a spin flip symmetry of the partition function, so an overall factor of 2 in the sum. The domain wall configurations must start and end at the same point due to periodic boundary conditions in time. Each domain wall contributes a weight $\frac{q}{q^2+1}$ per time step of the depth $2(t-1)$ lattice. Summing over multiple domain wall configurations, we have
\begin{equation}
\CF^{(2)}_{\rm RQC} = 2\bigg( 1 + c_1(n,t)\bigg(\frac{q}{q^2+1}\bigg)^{2(t-1)} + c_2(n,t)\bigg(\frac{q}{q^2+1}\bigg)^{4(t-1)} +\ldots\bigg)\,,
\end{equation}
where $c_i$ is the number of spin configurations with $i$ domain walls. For a single domain wall, there are at most $2(t-1)$ choose $t-1$ ways for the domain wall to random walk through the lattice to the same point, and $(n_g-1)$ starting points, where $n_g = \lfloor n/2\rfloor $ is the number of gates. Iterating up to configurations of $(n_g-1)$ domain walls, we may bound the frame potential as
\begin{align}
\CF^{(2)}_{\rm RQC} \leq 2\bigg( 1 + (n_g-1) \binom{2(t-1)}{t-1} \left(\frac{q}{q^2+1}\right)^{2(t-1)} + \binom{n_g-1}{2} \binom{2(t-1)}{t-1}^2 \left(\frac{q}{q^2+1}\right)^{4(t-1)}\nn + \ldots + \binom{n_g-1}{n_g-1 } \binom{2(t-1)}{t-1}^{n_g-1} \left(\frac{q}{q^2+1}\right)^{2(n_g-1)(t-1)} \bigg)\,,
\end{align}
with combinatorial factors enumerating multiple domain wall configurations and their associated weight. Continuing, we sum the binomial expansion in $n_g$
\begin{align}
\CF^{(2)}_{\rm RQC} &\leq 2 \bigg( 1 + \binom{2(t-1)}{t-1} \left(\frac{q}{q^2+1}\right)^{2(t-1)} \bigg)^{n_g-1}\nn
& < 2\bigg( 1 + \left(\frac{2 q}{q^2+1}\right)^{2(t-1)} \bigg)^{n_g-1}\,,
\label{eq:FP2bound}
\end{align}
upper bounding the binomial. We see that the second frame potential for random circuits decays as $\CF^{(2)}_{\rm RQC} \approx 2 ( 1 + 1/q^{2t})^n$, reaching its Haar value at a time scale $t \sim \log n$.

\vspace*{2pt}
Note that the bound is overcounting the domain wall configurations in two ways: we are considering a finite number of sites and the domain walls cannot cross spatial boundaries, and domain walls cannot cross, so the number of the paths that multiple domain walls can take are restricted. Lastly, we note that the function of $q$ in Eq.~\eqref{eq:FP2bound} is exponentially decaying in time for all local dimensions $q$, \ie $q=2$ and greater.

\subsubsection*{RQC 2-design time}
We now compute the 2-design time for random quantum circuits. The difference in frame potentials bounds the diamond distance between the channels as
\begin{equation}
\big\|\Phi^{(2)}_{\rm RQC} - \Phi^{(2)}_{\rm Haar}\big\|_\diamond^2 \leq d^{4} \big(\CF^{(2)}_{\rm RQC} - \CF^{(2)}_{\rm Haar}\big)\,.
\end{equation}
Using the upper bound in Eq.~\eqref{eq:FP2bound}, we can compute depth at which random circuits form an $\epsilon$-approximate 2-design from
\begin{equation}
q^{2n} \left( 2\bigg( 1+\bigg(\frac{2 q}{q^2+1}\bigg)^{2(t-1)} \bigg)^{n_g-1} -2 \right)^{1/2} \leq \epsilon\,.
\end{equation}
Taking the log of both sides, we expand out the binomial in the log and upper bound each term in terms of the leading order term, \ie the multidomain wall contributions are all bounded by the single domain wall term. After some manipulation, we find the circuit depth at which we form an $\epsilon$-approximate $2$-design to be
\begin{equation}
t_2 \geq C\big( 2n \log q + \log n + \log 1/\epsilon\big) \with C = \bigg(\log\frac{q^2+1}{2q}\bigg)^{-1}\,.
\label{eq:t2design}
\end{equation}
To get a sense of the scaling of $n$ with $q$, we see that for qubits with local dimension $q=2$ the linear term is $t_2 \sim 6.2 n$, and in the limit of large local dimension $q\ra \infty$ we find $t_2 \sim 2n$. 

Our analytic treatment of the second moment gives the 2-design time of $t_2 = O(n + \log 1/\epsilon)$, reproducing the known result in \cite{HL08,BHH12}. Given that the frame potential reaches its minimal value in $\log n$ time, we see from Eq.~\eqref{eq:t2design} that it takes an additional $O(n)$ time to get close in norm.

\subsubsection*{Counting domain walls}
Above we opted for a simple bound on the allowed domain wall configuration, but it is possible to compute the combinatorial coefficients exactly, accounting for the presence of boundaries and interactions between domain walls. The problem is equivalent to counting the number of non-intersection random walks on a finite 1D lattice. Consider two non-intersecting domain walls and for the moment ignore the effects of the boundaries. We can use the method of images \cite{Fisher84} to compute the number of non-intersecting configurations. For two random walkers starting at points $x$ and $y$, respectively, and ending at the same points after $2(t-1)$ time steps, there are at most $\binom{2(t-1)}{t-1}^2$ possible paths. To account for the crossings, note that any intersection of the two random walks from $x\ra x'$ and $y\ra y'$ can also be thought of as a path from $x\ra y'$ and $y\ra x'$. So the number of possible paths for two non-intersecting random walkers to return to their starting points after $2(t-1)$ time steps is
\begin{equation}
\binom{2(t-1)}{t-1}^2 - \binom{2(t-1)}{t-1-|x-y|}^2\,,
\end{equation}
where $|x-y|$ is the distance between the two walkers. Similarly, we can accommodate the effects of a boundary by again using a reflection. For a single random walker $x$ sites from the boundary returning to site $x$ after $2(t-1)$ steps, any path hitting the boundary can be thought of as a random walk from $-x$ to $x$. Thus the number of single domain wall configurations, accounting for both boundaries and summing over all starting points, is
\begin{equation}
c_1(n,t) = \sum_{x=1}^{n_g-1} \bigg(\binom{2(t-1)}{t-1} - \binom{2(t-1)}{t-1-2x} - \binom{2(t-1)}{t-1-2(n_g - x)}\bigg)\,.
\end{equation}
But this only takes into account paths crossing one of the boundaries any number of times. For very long times, we need to account for paths that hit the left boundary and then the right boundary, etc. We can again count the paths by repeating the method of images for multiple boundary intersections. Furthermore, we can iterate the same procedure to count multidomain wall configurations. For $p$ random walkers, one may consider a single random walk moving in $p$ dimensions, with a restriction on the coordinates from the non-intersection of the walkers. Iterating the method of images, one can determine the overall distribution of multiple random walks \cite{Fisher84,Huse84,Forrester91}. This will give an analytic expression for the multidomain wall contributions in the presence of boundaries.

\subsubsection*{Periodic boundary conditions}
If we instead start with a 1D array of qudits with periodic boundary conditions on the chain, it is straightforward to see that the above bound in Eq.~\eqref{eq:FP2bound} still holds: even if paths can wrap around the circuit, there are still at most $2^{2(t-1)}$ paths for a single domain wall after $2(t-1)$ time steps. Moreover, periodic spatial boundary conditions would mean that configurations with an odd number of domain walls would be disallowed. The leading order contribution would arise from double domain wall configurations and thus the frame potential would still decay in $\log n$ time, albeit slightly quicker than with open boundary conditions.

\section{\ktitle-designs from statistical mechanics}
\label{sec:kdesigns}
We now proceed to discussing higher moments of random circuits. The statistical mapping discussed in Sec.~\ref{sec:statmech} allows us to write the $k$-th frame potential for random quantum circuits as a classical partition function of spins $\sigma\in S_k$ on a triangular lattice of width $n_g$ and depth $2(t-1)$ with periodic boundary conditions in time. In the second moment, the exited states were spin configurations with one type of domain wall separating regions of identity and swap permutations. In this sense the domain wall represented a swap, passing through the domain wall took us from $\iden$ spins to $S$ spins. Moreover, the domain walls could not intersect, pair create, or annihilate, and their counting was amenable to a simple combinatorial treatment.

At higher $k$, this simple picture no longer holds. We have multiple different types of domain walls which can interact nontrivially, including interactions where domain walls end. Recall that the plaquette terms $J^{\sigma_1}_{\sigma_2 \sigma_3} $ are functions of three permutations $\sigma \in S_k$, defined as a sum of Weingarten functions and permutation inner products over an internal permutation as 
\begin{equation}
J^{\sigma_1}_{\sigma_2 \sigma_3} = \sum_{\tau \in S_k} \Wg(\sigma_1^{-1}\tau,q^2) q^{\ell(\sigma_2^{-1}\tau)}q^{\ell(\sigma_3^{-1}\tau)}\,.
\end{equation}
Computing these quantities for all $\sigma\in S_k$ will tell us the allowed terms in the $k$-th partition function. The types of plaquette terms can be understood in terms of domain walls between the different permutations. As in \cite{RQCstatmech}, domain walls for the $k$-th moment will denote transpositions on $k$ elements, a generating set for the symmetric group $S_k$. For instance, in the third moment, if we have neighboring spins $\{1,2,3\}$ and $\{1,3,2\}$, we say they are separated by a domain wall representing the $(23)$ transposition. In general, there are $\binom{k}{2}$ transpositions and any two elements of $S_k$ differ by at most $k-1$ transpositions. Even for $k=3$, we find additional complications as there is an interaction term with 4 incoming and 2 outgoing domain walls. We prove some general properties of domain walls in App.~\ref{app:kplaq} and give the explicit weights of plaquette terms for the first few moments.

\subsection{Lattice domain wall configurations}

\subsubsection*{Single domain wall sector}
For general $k$, there are significant complications in the multi-domain wall contributions to the partition function. But as we will see, the single domain wall sector remains simple. This stems from two facts about the plaquette terms for general $k$, which we prove in App.~\ref{app:kplaq}. The first is that for spins $\sigma\in S_k$, we have $J^{\sigma}_{\sigma\sigma} = 1$ when all spins are the same, and that $J^\sigma_{\sigma'\sigma'} = 0$ for $\sigma\neq \sigma'$, \ie the weight for no domain walls is just one and interactions with just two domain walls annhilating are disallowed. This means we cannot have closed loops in the circuit. These statements both arise from the fact that contracting a $U$ with a $U^\dagger$ in the average gives the identity.

The other fact is that we cannot annihilate incoming domain walls in plaquettes with only one outgoing domain wall. This arises from the statement that if $\sigma_2$ and $\sigma_3$ in  $J^{\sigma_1}_{\sigma_2\sigma_3}$ only differ by a single transposition, then the resulting Haar integral reduces to a second moment calculation, which can only give an identity or swap. This translates to only allowing one incoming domain wall if there is only one outgoing domain wall, and forbids domain wall annihilation in the single domain wall sector. More generally, incoming domain walls can annhiliate in plaquettes with multiple outgoing domain walls, but these properties guarantee the independence of the single domain wall sector. 

We now count the configurations of single domain walls. A domain wall corresponds to a single transposition, of which there are $\binom{k}{2}$ for spins $\sigma\in S_k$. Accounting for the $k!$ spin flip symmetry of the lattice, we then count the directed random walks through the depth $2(t-1)$ lattice, with $(n_g-1)$ starting points for a single domain wall and assigning the weight, we find
\begin{equation}
\sum_{1~{\rm dw}} wt(q,t) \leq (n_g-1)\binom{k}{2}\binom{2(t-1)}{t-1} \Big(\frac{q}{q^2+1}\Big)^{2(t-1)}\,.
\end{equation}

\begin{figure}
\centering
\begin{tikzpicture}[thick]
\node at (0,0) {\includegraphics[width=3.5cm]{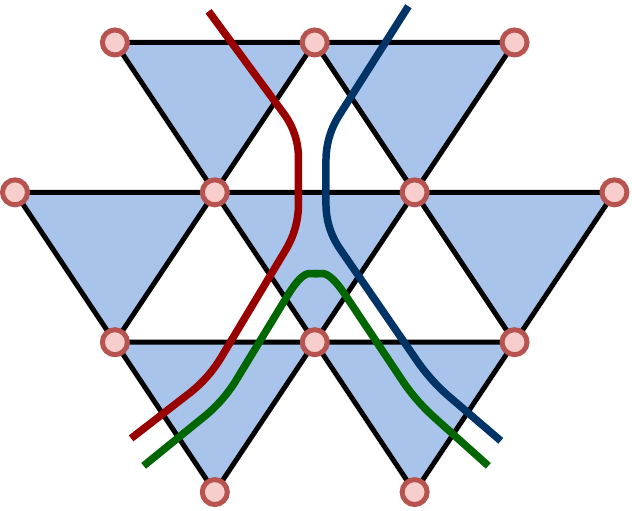}};
\draw[thick,->] (2,-1.2) -- (2,0.8) node[anchor=south] {$t$};
\end{tikzpicture}
\caption{An example of a domain wall annihilating at a vertex, creating a closed loop in the circuit. The domain walls represent transpositions, generators of $S_k$, which we denoted with different colors.}
\label{fig:k3dw}
\end{figure}

\subsubsection*{Multi-domain wall sector}
For the $k$ -th moment of the random circuit, we will have contributions from domain walls which can interact and can pair create and annihilate. For instance, at $k=3$ we already have an interaction vertex with 4 ingoing domain walls and 2 outgoing, which permits spin configurations with closed loops, as shown in Fig.~\ref{fig:k3dw}. This means we cannot separate the contributions in the partition function into sectors of differing numbers of domain walls and simply count them. 

We gave a bound on single domain wall configurations above. Similarly, for double domain wall configurations with no intersections, we have the bound
\begin{equation}
< (n_g-1)^2\binom{k}{2}^2\binom{2(t-1)}{t-1}^2 \Big(\frac{q}{q^2+1}\Big)^{4(t-1)}\,.
\end{equation}
There are at most $2(n_g-1)(k-1)$ domain walls at $t=0$, as any permutation in $S_k$ can be reached with at most $k-1$ transpositions. For $\sim nk$ multidomain wall configurations with no intersections between domain walls, all can be bounded by powers of the single domain wall contribution.

The technical hurdle that arises for higher moments is in dealing with possible intersections. Interaction terms between multiple different types of domain walls are allowed with weights depending on the types of incoming domain walls. Some of the interaction terms up to $k=4$ are given in App.~\ref{app:kplaq}, but we have computed all such terms up to $k=6$. Even for two domain walls with a single intersection, we may then have many configurations of multiple closed loops in the lattice. To rigorously bound their contribution to the partition function we must bound the number of all such possible configurations. Nevertheless, these interactions are suppressed by many factors of the local dimension, seemingly ensuring the dominance of the single domain wall sector.

\subsubsection*{RQC \ktitle-design time}
Given what we have understood so far, we can sketch how the $k$-design time should arise from the $k$-th frame potential. We have the contribution from the ground states and single domain wall sector, plus higher order contributions
\begin{equation}
\CF^{(k)}_{\rm RQC} = k!\bigg( 1+ (n_g-1)\binom{k}{2}\binom{2(t-1)}{t-1}\Big(\frac{q}{q^2+1}\Big)^{2(t-1)}+\ldots\bigg)\,.
\label{eq:FPgenk}
\end{equation}
Recalling how the frame potentials bound the distance between $k$-fold channels,
\begin{equation}
\big\|\Phi^{(k)}_{\rm RQC} - \Phi^{(k)}_{\rm Haar}\big\|_\diamond^2 \leq d^{2k} \big(\CF^{(k)}_{\rm RQC} - \CF^{(k)}_{\rm Haar}\big)\,,
\end{equation}
we see that the single domain wall sector will give the depth at which we form an approximate $k$-design to be 
$t_k \sim nk \log q + k\log k+ \log(1/\epsilon)$, up to constant factors. For moments $k\leq q^n$, we find a depth $O(nk)$.

\subsection{Designs from single domain walls}

\subsubsection*{\ktitle-designs at large $q$}
The analysis for general moments becomes simple in the limit of large local dimension $q$. We prove in App.~\ref{app:kplaq} that the only plaquette terms that contribute at order $1/q$ are single domain walls. More generally, interaction terms with $l$ outgoing domain walls will contribute at most at order $1/q^l$, with additional penalties by factors of $1/q$ if we have domain walls annihilating in an interaction term. So more complicated interactions are generally greatly suppressed in $q$, with only single freely propagating domain walls contributing at leading order in $1/q$. Thus in the $q\ra \infty$ limit we can bound the higher order domain wall terms in Eq.~\eqref{eq:FPgenk} and find the difference in the $k$-th frame potentials to be
\begin{equation}
\CF^{(k)}_{\rm RQC} - \CF^{(k)}_{\rm Haar} \leq k! \bigg(nk^2 \bigg(\frac{2}{q}\bigg)^{2(t-1)} + O\bigg(\frac{1}{q^{4t}}\bigg)\bigg)\,,
\end{equation}
from which we recover the $k$-design time in the large $q$ limit
\begin{equation}
t_k \geq C \big(2 n k \log q + k\log k + \log (nk^2) +\log(1/\epsilon) \big)\,,
\end{equation}
where $C = \log^{-1} (q/2)$.

\subsubsection*{\ktitle-designs at finite $q$}
For a given $k$, we can explicitly compute the plaquette terms as functions of $q$, but the number of unique interaction terms grows rapidly with $k$. We can understand their behavior to leading order in $1/q$, but it is unclear how much the exact functions matter in trying to bound the higher order contributions to the $k$-th partition function, or if additional subtleties arise for exponentially large moments and small $q$. 

One issue that arises when considering higher moments is that the Weingarten functions have poles in $q^2$ at $\{\pm 1, \ldots, \pm (k-1)\}$. Thus, one might worry about the partition function for $k>q^2$, there are a few reasons this is not an issue, which we discuss in more detail in App.~\ref{app:kplaq}. For moments greater than the dimension, the Weingarten function is modified (restricting the sum over integer partitions) to free the denominator of zeroes. But we can drop this restriction and will find that the poles in the plaquette terms will cancel for any physical value of the dimension $q\in \mathbb{N}$. Moreover, the plaquette rules for general $k$ we derive in App.~\ref{app:kplaq} arise from properties of permutations acting on the $k$-fold space and hold for any $q$. 

We end this section on a slightly more speculative note. Taking the limit of large local dimension ensures the dominance of the single domain wall terms, but such a limit is likely not necessary. Configurations with multiple domain walls decay rapidly, and the more complicated the types of interactions, the stronger the suppression. Furthermore, the interaction terms with annihilating domain walls have negative weight, further reducing their contribution to the partition function. Thus, we conjecture that the single domain wall terms dominate for general $k$.

\begin{conjecture} The single domain wall sector of the lattice partition function dominates the multidomain wall sectors for higher moments $k$ and any local dimension $q$. 
\end{conjecture}

Lastly, a lower bound on the depth is known (Prop. 8 in \cite{BHH12}). For $\epsilon\leq1/4$ and $k\leq d^{1/2}$, an $\epsilon$-approximate $k$-design on $n$ qudits of local dimension $q$ must have circuit depth at least
\begin{equation}
t_k\geq \frac{nk}{5q^4 \log(nk)}\,.
\end{equation}
If the single domain wall contribution individually bounds each of the $\sim nk$ domain wall terms, then the depth at which random circuits form an approximate $k$-design is $O(nk)$, with a dependence on $n$ and $k$ that is essentially optimal.

\section{Discussion}
We studied the convergence of random quantum circuits to unitary $k$-designs and by exactly expressing the frame potential as a lattice partition function, were able to give an analytic treatment of the second moment, showing that random circuits form approximate 2-designs in $O(n)$ depth. By then proving some properties of the partition function for the $k$-th moment, we then showed that in large $q$ limit random circuits form $k$-designs in $O(nk)$ depth. We further conjectured that the terms corresponding to single domain walls in the spin system bound the multi-domain wall contributions for any local dimension $q$, from which it holds that the $k$-design time is $O(nk)$ and thus random circuits are optimal implementations of randomness. A rigorous bound on the higher order terms in the $k$-th partition function is needed to make this precise, and we hope to return to this in future work.

As we have seen, the statistical mechanical mapping renders certain random circuit calculations analytically tractable. It would be interesting to see if these techniques could be applied to the higher-dimensional random circuits considered in \cite{HM18}. Moreover, we focused on circuits with a fixed geometry, which gave a partition function on a regular triangular lattice. Considering random circuits with gates applied randomly to neighboring pairs of qudits should give the partition function of a spin system on random triangular tilings of the strip. Furthermore, there has been recent interest in other random circuit models with conservation laws \cite{KVH17,Rakovszky17}, local symmetries including time-reversal symmetry \cite{RQCsym}, and Floquet random circuits \cite{FRQC17,FRQC18}. It would be interesting to see if lattice models could be defined for each of these circuits. For instance, one could define a lattice mapping for the circuits considered in \cite{RQCsym} using Weingarten calculus for the different symmetry groups and show that circuits built from random orthogonal gates converge to the Haar measure on $O(d)$. 
It would also be interesting to investigate whether the charge conserving random circuits in \cite{KVH17,Rakovszky17} form $k$-designs within fixed charge sectors.

\subsubsection*{Acknowledgments}
The author would like to thank Fernando Brand\~ao,  Jacob Bridgeman, Richard Kueng, Zi-Wen Liu, Saeed Mehraban, Beni Yoshida, and especially Adam Nahum for helpful discussions, as well as \'Alvaro Alhambra and Thom Bohdanowicz for discussions and comments on the draft. The author also thanks the IQIM at Caltech, McGill University, and UC Berkeley for hospitality during the completion of part of this work. Research at Perimeter Institute is supported by the Government of Canada through the Department of Innovation, Science and Economic Development Canada and by the Province of Ontario through the Ministry of Research, Innovation and Science.

\appendix

\section{Approximate designs and random unitaries}
\label{app:designs}

In this appendix we quickly review some definitions related to approximate $k$-designs and Haar integration. We introduced the notion of an approximate $k$-design in Def.~\ref{def:kdesign} as the distance of the $k$-fold channels in diamond norm. After providing a few relevant definitions we will also show how to bound the diamond norm in terms of the frame potential. 

\subsubsection*{\ktitle-fold channels}
The $k$-fold channel of an operator $\op$ with respect to an ensemble $\CE$ of unitaries is defined as
\begin{equation}
\Phi^{(k)}_\CE(\op) \equiv \int dU\, U^{\otimes k}(\op) U^\dagger{}^{\otimes k}\,,
\end{equation}
here written for a continuous ensemble. As we discussed, an {\it exact unitary $k$-design} is an ensemble $\CE$, a subset of the unitary group equipped with some probability measure, for which the $k$-fold channels are equal $\Phi^{(k)}_\CE(\op) = \Phi^{(k)}_{\rm Haar} (\op)$, for all operators acting on the $k$-fold Hilbert space $\op \in \CA(\CH^{\otimes k})$. 

It will also be convenient to introduce the moment operator for an ensemble $\CE$, closely related to the $k$-fold channel. The $k$-th moment operator $\widehat \Phi^{(k)}_\CE$ is defined as
\begin{equation}
\widehat\Phi^{(k)}_\CE \equiv \int dU\, U^{\otimes k}\otimes U^\dagger{}^{\otimes k}\,.
\end{equation}
The $k$-th moment operators for an ensemble $\CE$ and the Haar ensemble are also equal $\widehat\Phi^{(k)}_\CE = \widehat\Phi^{(k)}_{\rm Haar}$ if and only if $\CE$ forms a $k$-design.

\subsubsection*{Operator norms}
The Schatten $p$-norm of an operator $\op$ is defined for $p\geq 1$ as $\|\op \|_p \equiv \big(\Tr |\op|^p\big)^{1/p}$, which obeys monotonicity of norms $\|\op\|_q\leq \|\op\|_p$ for $q\geq p$. The superoperator norm of a quantum channel is defined for any $p\geq 1$ as
\begin{equation}
\|\Phi\|_{p\ra p} \equiv \sup_{\op \neq 0} \frac{\|\Phi(\op)\|_p}{\|\op\|_p}\,.
\end{equation}
The diamond norm of a quantum channel $\Phi$ is defined as
\begin{equation}
\|\Phi\|_\diamond \equiv \sup_d \|\Phi \otimes {\rm Id}_d \|_{1\ra 1}\,,
\end{equation}
where ${\rm Id}_d$ is the identity channel on a $d$-dimensional ancilla. The diamond norm captures the distinguishability of quantum channels. One relation between the diamond norm and the operator 2-norm we will need is $\|\Phi\|_\diamond \leq d^k \|\Phi\|_{2\ra 2} $, for $\Phi$ acting on $k$-fold operators. 

\subsubsection*{Approximate $k$-design}
We define an $\epsilon$-approximate $k$-design as an ensemble of unitaries $\CE$ such that the $k$-fold channel with respect to $\CE$ is
\begin{equation}
\big\| \Phi^{(k)}_\CE - \Phi^{(k)}_{\rm Haar} \big\|_\diamond \leq \epsilon\,,
\end{equation}
meaning that the ensembles are close with respect to the diamond norm. The notion of an approximate design makes precise a distance to randomness, capturing how close an ensemble is to replicating moments of $U(d)$. There are other definitions of approximate designs, some employing different norms, which bound each other up to factors of the dimension, as is reviewed in \cite{LowThesis}.

\subsubsection*{Frame potential}
In general, diamond norms are difficult to compute. There are other quantities which capture the distance of an ensemble to the Haar ensemble which are more tractable. The frame potential is defined as a double average over an ensemble of unitaries $\CE$ as \cite{Gross07,Scott08}
\begin{equation}
\CF^{(k)}_\CE = \int_{U,V\in \CE} dUdV\, \big| \Tr(U^\dagger V)\big|^{2k}\,.
\end{equation}
The frame potential is related to the 2-norm distance of the ensembles and is lower bounded by its Haar value
\begin{equation}
\CF^{(k)}_\CE \geq \CF^{(k)}_{\rm Haar} \where \CF^{(k)}_{\rm Haar} = k!\,,
\end{equation}
for $d\geq k$. In this work, we restrict ourselves to studying moments of random circuits $k<d=q^n$, but note that the Haar value is modified once we probe moments greater than the dimension. The Haar value of the frame potential arises from an old result on the moments of traces of random unitaries \cite{Diaconis94}. When $k>d$, then $|\Tr(U)|^{2k}$ is equal to the number of permutations in $S_k$ with the longest increasing subsequence of length $\leq d$ \cite{Rains98}. 

The difference in frame potentials is equal to the 2-norm distance between the moment operators as follows. Consider the operator
\begin{equation}
\Delta \equiv \widehat\Phi^{(k)}_\CE -\widehat\Phi^{(k)}_{\rm Haar} = \int_\CE dU\, U^{\otimes k}\otimes U^\dagger{}^{\otimes k} -  \int_{\rm Haar} dU\, U^{\otimes k}\otimes U^\dagger{}^{\otimes k}\,,
\end{equation}
and note that
\begin{equation}
\Tr(\Delta^\dagger \Delta) = \CF^{(k)}_\CE - 2 \CF^{(k)}_{\rm Haar} + \CF^{(k)}_{\rm Haar} = \CF^{(k)}_\CE - \CF^{(k)}_{\rm Haar}\,,
\end{equation}
where in the middle term we used the left/right invariance of the Haar measure to absorb the terms from the $\CE$ average, giving the frame potential for the Haar ensemble. Moreover, as $\Tr(\Delta^\dagger \Delta)\geq 0$, this shows that the frame potential for any ensemble of unitaries $\CE$, is lower bounded by the Haar value $\CF^{(k)}_\CE \geq \CF^{(k)}_{\rm Haar}$ as we claimed above. We now have that
\begin{equation}
\big\|\widehat\Phi^{(k)}_\CE-\widehat\Phi^{(k)}_{\rm Haar}\big\|_2^2 = \CF^{(k)}_\CE-\CF^{(k)}_{\rm Haar}\,.
\end{equation}

As such, we can bound the diamond norm of the difference in $k$-fold channels in terms of the frame potentials of the two ensembles noting that
\begin{equation}
\big\| \Phi^{(k)}_\CE - \Phi^{(k)}_{\rm Haar} \big\|_{\diamond} \leq d^k \big\| \Phi^{(k)}_\CE - \Phi^{(k)}_{\rm Haar} \big\|_{2\ra 2} = d^k \big\| \widehat\Phi^{(k)}_\CE - \widehat\Phi^{(k)}_{\rm Haar} \big\|_\infty \leq d^k \big\| \widehat\Phi^{(k)}_\CE - \widehat\Phi^{(k)}_{\rm Haar} \big\|_2\,,
\end{equation}
and therefore
\begin{equation}
\big\| \Phi^{(k)}_\CE - \Phi^{(k)}_{\rm Haar} \big\|_{\diamond}^2 \leq d^{2k} \big(\CF^{(k)}_\CE-\CF^{(k)}_{\rm Haar}\big)\,.
\end{equation}
Note that had we defined approximate designs in terms of the trace norm of the moment operators, the factors of $d$ to bound in terms of the frame potential would be the same. 

\subsubsection*{Haar integration and Weingarten calculus}
The general expression for integrating the $k$-th moment of $U(d)$ is \cite{Collins02,Collins04}
\begin{equation}
\int dU\, U_{i_1j_1}\ldots U_{i_k j_k} U^\dagger_{\ell_1 m_1}\ldots U^\dagger_{\ell_k m_k} = \sum_{\sigma,\tau\in S_k} \delta_\sigma (\vec \imath\,| \vec m) \delta_\tau (\vec \jmath\,| \vec\ell\, ) \Wg(\sigma^{-1}\tau,d)\,,
\label{eq:Haarint}
\end{equation}
where we sum over elements of the permutation group $S_k$ and define a $\delta$-function contraction indexed by a permutation $\sigma$ as $\delta_\sigma(\vec \imath\,|\vec\jmath\,) = \delta_{i_1, j_{\sigma(1)}} \ldots \delta_{i_k, j_{\sigma(k)}}$. The Weingarten function is a function of a permutation $\sigma\in S_k$ and admits an expansion in terms of characters of the symmetric group as
\begin{equation}
\Wg(\sigma, d) = \frac{1}{k!} \sum_{\lambda \vdash k} \frac{\chi_\lambda(\sigma) f_\lambda}{c_\lambda(d)}\,, \where c_\lambda(d) = \prod_{(i,j)\in \lambda} (d+j-1)
\label{eq:Wg}
\end{equation}
and we sum over integer partitions of $k$, labelling the irreducible representations of $S_k$, $\chi_\lambda(\sigma)$ is an irreducible character of $\lambda$, and $f_\lambda$ is the dimension of the $\lambda$ irrep. The polynomial $c_\lambda(d)$ is a product over the coordinates $(i,j)$ of the Young diagram of the irrep $\lambda$. The Weingarten functions are equivalently computed as the matrix inverse of the Gram matrix of permutation operators on the $k$-fold space. 

Strictly speaking, the expression for the Weingarten function in Eq.~\ref{eq:Wg} is valid for any $d\geq k$, such that the denominator is free of zeroes. For $k>d$, the expression is modified by taking the sum instead over integer partitions of length $\ell(\lambda)\leq d$. But we may drop this restriction and use Eq.~\ref{eq:Wg} for general moments (Prop 2.5 in \cite{Collins04}). $\Wg(\sigma,d)$ is a rational function of $d$ with finitely many poles, but the expression for Haar integration in Eq.~\eqref{eq:Haarint} will remain true for any $d\in \mathbb{N}$ using the full expression for the Weingarten function after some cancellation of poles.

\section{Plaquette terms for higher \ktitle}
\label{app:kplaq}
In Sec.~\ref{sec:2design}, we explicitly derived the plaquette terms for $k=2$, which we could interpret as simple rules regarding the domain wall configurations separating regions of $\iden$'s and $S$'s in the lattice model. But in order to discuss the nontrivial spin configurations contributing to the $\CF^{(k)}_{\rm RQC}$ partition function, we need to understand the properties of plaquette terms for general $k$. 

In this appendix, we will denote the plaquette terms with time running up, such that the triangles are downward pointing as
\begin{equation}
J^{\sigma_1}_{\sigma_2\sigma_3} =
\begin{tikzpicture}[thick,scale=0.5,baseline=0.3cm]
\draw (0,0) -- (1,1.6) -- (-1,1.6) -- (0,0);
\node[anchor=north] at (0,0.1) {$\sigma_1$};
\node[anchor=west] at (0.9,1.5) {$\sigma_3$};
\node[anchor=east] at (-0.9,1.5) {$\sigma_2$};
\end{tikzpicture}
= \sum_{\tau\in S_k} \Wg(\sigma_1^{-1}\tau,q^2) q^{\ell(\tau^{-1} \sigma_2)}q^{\ell(\tau^{-1} \sigma_3)}\,,
\end{equation}
where again $\ell(\tau^{-1} \sigma)$ denotes the length of the cycle-type of the permutation product; equivalent to the number of closed loops in the product. For instance, if $\tau = \sigma^{-1}$, then $\ell(\iden) = k$ as the cycle-type of the identity is the partition into all ones. 

From now on in this appendix we will denote the plaquettes only in terms of the domain walls representing transpositions between the permutations and drop the explicit dependence on permutations. For instance, the empty plaquette $\begin{tikzpicture}[thick,scale=0.25,baseline=0.1cm]
\draw (0,0) -- (1,1.6) -- (-1,1.6) -- (0,0);
\end{tikzpicture}$ corresponds to $J^{\sigma}_{\sigma\sigma}$, \ie the permutations differ by no transpositions. The plaquette
$\begin{tikzpicture}[thick,scale=0.25,baseline=0.1cm]
\draw (0,0) -- (1,1.6) -- (-1,1.6) -- (0,0);
\draw[blue] (-0.9,0.2) -- (-0.2,1) -- (-0.2,2);
\draw[red] (0.9,0.2) -- (0.2,1) -- (0.2,2);
\end{tikzpicture}$ corresponds to $J^{\sigma_1}_{\sigma_2\sigma_3}$, with $\sigma_1$ differing from $\sigma_2$ and $\sigma_3$ by a single transposition, and $\sigma_2$ and $\sigma_3$ differing by two transpositions. 

Now we will prove some general properties of the plaquette terms for any $k$, which were pointed out in \cite{RQCstatmech}. First, we want to show that plaquette terms evaluated on the same permutation $J^{\sigma}_{\sigma\sigma} =1$ for $\sigma\in S_k$ and $J^{\sigma}_{\sigma'\sigma'}=0 $ when $\sigma\neq \sigma'$. Recalling how the plaquette terms arise from the moments of a single 2-site unitary $U^{\otimes k} \otimes U^\dagger{}^{\otimes k}$, the permutations $\sigma_2$ and $\sigma_3$ act on the outgoing indices of $U$ and ingoing indices of $U^\dagger$ as
\begin{equation}
\begin{tikzpicture}[baseline,scale=0.55]
\node at (-2.4,0) {$U^{\otimes k}$};
\draw[rounded corners] (-3.3,-1) rectangle (-1.5,1);
\draw[thick,rounded corners] (-1.5,0.5) -- (1.5,0.5);
\draw[thick,rounded corners] (-1.5,-0.5) -- (1.5,-0.5);
\draw[rounded corners,fill=white] (-0.5,0.1) rectangle (0.5,1);
\draw[rounded corners,fill=white] (-0.5,-1) rectangle (0.5,-0.1);
\node at (0,0.5) {{\small $\sigma_2$}};
\node at (0,-0.6) {{\small $\sigma_3$}};
\node at (2.4,0) {$U^\dagger{}^{\otimes k}$};
\draw[rounded corners] (1.5,-1) rectangle (3.3,1);
\foreach \x in {-4.1,3.3}
{\draw[thick] (\x,0.5) -- (\x+0.8,0.5);
\draw[thick] (\x,-0.5) -- (\x+0.8,-0.5);}
\end{tikzpicture}
\label{eq:Ugatek}
\end{equation}
Taking $\sigma_2 = \sigma = \sigma_3 $, we are simply contracting $U$'s with $U^\dagger$'s, which cancel and just give the permutation operator on the $k$-fold space $P_\sigma$.\footnote{A permutation operator $P_\sigma$ permutes the $k$ copies of $\CH^{\otimes k}$ and acts on computational basis elements as $P_\sigma \ket{i_1\ldots i_k} = \ket{i_{\sigma^{-1}(1)}\ldots i_{\sigma^{-1}(k)}}$.} Therefore, plaquette terms with all the same permutation $\sigma\in S_k$ have weight one, and plaquettes with any number of ingoing domain walls and no outgoing domain walls have weight zero
\begin{equation}
\begin{tikzpicture}[thick,scale=0.5,baseline=0.3cm]
\draw (0,0) -- (1,1.6) -- (-1,1.6) -- (0,0);
\end{tikzpicture} = 1 ~\and ~
\begin{tikzpicture}[thick,scale=0.5,baseline=0.3cm]
\draw (0,0) -- (1,1.6) -- (-1,1.6) -- (0,0);
\draw[red] (-0.7,0.1) -- (0,0.9) -- (0.7,0.1);
\end{tikzpicture} = 0\,,
\end{equation}
meaning that domain walls cannot simply annihilate. The same statement holds true for any number of incoming domain walls and no outgoing domain walls.

Next, consider Eq.~\eqref{eq:Ugatek} with $\sigma_2$ and $\sigma_3$ differing by a single transposition, corresponding to a single outgoing domain wall. This means that the $k$-fold unitaries $U^{\otimes k}$ are contracted with their daggered counterparts and cancel on all but two of the $k$-fold unitaries, and we find 
\begin{equation}
\begin{tikzpicture}[scale=0.6,baseline=-0.1cm]
\node[anchor=south] at (-2,-1.15) {$U$};
\node[anchor=south] at (-2,0.25) {$U$};
\node[anchor=south] at (2.05,0.25) {$U^\dagger$};
\node[anchor=south] at (2.05,-1.15) {$U^\dagger$};
\node at (0,0.7) {$\iden$};
\node at (0,-0.7) {$S$};
\foreach \x in {-2.5,-0.5,1.5}
{\draw[rounded corners] (\x,0.2) rectangle (\x+1,1.2);
\draw[rounded corners] (\x,-0.2) rectangle (\x+1,-1.2);}
\foreach \x in {-1.5,0.5}
{\draw[thick] (\x,0.9) -- (\x+1,0.9);
\draw[thick] (\x,-0.9) -- (\x+1,-0.9);}
\foreach \y in {0.9,0.5,-0.5,-0.9}
{\draw[thick] (-2.5,\y) -- (-3.1,\y);
\draw[thick] (2.5,\y) -- (3.1,\y);}
\draw[thick,rounded corners] (-0.5,0.5) -- (-0.8,0.5) -- (-1.2,-0.5) -- (-1.5,-0.5);
\draw[thick,rounded corners] (-0.5,-0.5) -- (-0.8,-0.5) -- (-1.2,0.5) -- (-1.5,0.5);
\draw[thick,rounded corners] (0.5,0.5) -- (0.8,0.5) -- (1.2,-0.5) -- (1.5,-0.5);
\draw[thick,rounded corners] (0.5,-0.5) -- (0.8,-0.5) -- (1.2,0.5) -- (1.5,0.5);
\end{tikzpicture}
= \frac{q}{q^2+1} P_{\iden} + \frac{q}{q^2+1} P_{S}\,,
\end{equation}
on the two factors of the $k$-fold space $\CH^{\otimes k}$ for which $\sigma_2$ and $\sigma_3$ differ by a transposition. Computing the second moment above, we find that result is an identity or swap on the two factors of $\CH^{\otimes k}$ for which $\sigma_2$ and $\sigma_3$ differ, \ie $\sigma_1$ can only differ from either $\sigma_2$ or $\sigma_3$ by a transposition. Thus we find that plaquettes with only one outgoing domain wall must have only one ingoing domain wall, expressed diagrammatically as
\begin{equation}
\begin{tikzpicture}[thick,scale=0.5,baseline=0.3cm]
\draw (0,0) -- (1,1.6) -- (-1,1.6) -- (0,0);
\draw[red] (-0.8,0.2) -- (0,1) -- (0,2);
\end{tikzpicture} =\begin{tikzpicture}[thick,scale=0.5,baseline=0.3cm]
\draw (0,0) -- (1,1.6) -- (-1,1.6) -- (0,0);
\draw[red] (0.8,0.2) -- (0,1) -- (0,2);
\end{tikzpicture} = \frac{q^2}{q^2+1} ~\and~
\begin{tikzpicture}[thick,scale=0.5,baseline=0.3cm]
\draw (0,0) -- (1,1.6) -- (-1,1.6) -- (0,0);
\draw[blue] (-0.8,0.2) -- (-0.08,1.03) -- (-0.08,2);
\draw[red] (-0.7,0.1) -- (0,0.9) -- (0.7,0.1);
\end{tikzpicture} = \begin{tikzpicture}[thick,scale=0.5,baseline=0.3cm]
\draw (0,0) -- (1,1.6) -- (-1,1.6) -- (0,0);
\draw[blue] (0.8,0.2) -- (0.08,1.03) -- (0.08,2);
\draw[red] (-0.7,0.1) -- (0,0.9) -- (0.7,0.1);
\end{tikzpicture} =0\,.
\end{equation}

We can also arrive at the same statements about the plaquette terms for general $k$ from properties of the Weingarten function \cite{RQCstatmech}, where $\Wg$'s are the elements of the inverse of the matrix of inner products of permutation operators. But this approach become subtle for $k>d$, when the matrix inverse is not strictly well-defined.

Lastly, we note that the above discussion makes it clear that the plaquette rules for do not actually depend on $k$, but only on the factors for which $\sigma_2$ and $\sigma_3$ in Eq.~\eqref{eq:Ugatek} differ, which reduces to a moment calculation on those factors. For instance, if $\sigma_2$ and $\sigma_3$ differ by $l$ transpositions, then the calculation of the  $J^{\sigma_1}_{\sigma_2\sigma_3}$'s will only involve computations of at most the $2l$-th moment, and possibly less if the transpositions overlap. 

\subsubsection*{Plaquettes at large $q$}
Now we discuss the weights in the limit of large local dimension and show that an interaction term with $l$ ingoing domain walls contributes at most at order $\sim 1/q^l$. For the $k$-th moment, Weingarten functions have following asymptotic expansion in the dimension $q^2$ \cite{Weingarten78}
\begin{equation}
\Wg(\sigma,q^2) \sim \frac{1}{q^{2(2k-\ell(\sigma))}}\,,
\end{equation}
with the sign determined by ${\rm sgn}(\sigma)$. Precise expressions and bounds for the asymptotic form of $\Wg$ were recently given in \cite{CollinsMat17}. 
As the length of the cycle type is at most $k$, the $\Wg$'s for $\sigma\in S_k$ contribute at order $1/q^{2k}$ to $1/q^{4k}$. The longer the length of the cycle-type of $\sigma$, the higher order the contribution. At leading order in $1/q$, the interaction term $J$ can be written
\begin{equation}
J^{\sigma_1}_{\sigma_2 \sigma_3} \sim \sum_{\tau\in S_k} \frac{q^{\ell(\tau^{-1}\sigma_2)} q^{\ell(\tau^{-1}\sigma_3)}}{q^{2(2k-\ell(\sigma_1^{-1}\tau))}}\,.
\end{equation}
The length of the cycle type of $\sigma$ is related to the number of transpositions as $\ell(\sigma) = k-|\sigma|$, where $|\sigma|$ is the minimal number of transpositions needed to write $\sigma$. The expression in the sum simply becomes $(q^{2|\sigma_1^{-1}\tau|} q^{|\sigma_2^{-1}\tau|}  q^{|\sigma_3^{-1}\tau|})^{-1}$.

Given an interaction term with $l$ ingoing domain walls, the term contributing at leading order in the sum is $\tau = \sigma_1$, corresponding to the leading Weingarten function. As $\sigma_1$ differs from $\sigma_2$ and $\sigma_3$ by a total of $l$ transpositions, the interaction contributes at order $\sim 1/q^{l}$. For an interaction term with $l$ ingoing and $l$ outgoing domain walls, and thus no annihilating domain walls, this term is the only term contributing at leading order in $1/q$. But if we have domain wall annihilation in an interaction, other terms in the sum contribute at the same leading order. Take the simplest interaction plaquette of this sort, with 4 ingoing and 2 outgoing domain walls. The $\tau = \sigma_1$ term contributes $\sim 1/q^4$, but there is also a $\tau$ which differs from each permutation by a single transposition, thus also contributing at $\sim 1/q^4$ but involving the next-to-leading order Weingarten function. As the sign of $\Wg(\sigma,d)$ depends on the the signature of the permutation, plaquettes with annhilating domain walls can have negative weight.

\subsubsection*{Explicit plaquette weights}
Below we list some of the nontrivial plaquette terms for higher $k$, but we have computed all terms up to $k=6$. Note that as the plaquette terms are symmetric in the outgoing permutations, all the triangles weights below are reflection symmetric about the vertical axis.

\smallskip
\ni For $k=2$, the only we only have a single domain wall corresponding to the swap transposition
\begin{equation}
\begin{tikzpicture}[thick,scale=0.5,baseline=0.3cm]
\draw (0,0) -- (1,1.6) -- (-1,1.6) -- (0,0);
\draw[red] (-0.8,0.2) -- (0,1) -- (0,2);
\end{tikzpicture}
= \frac{q}{q^2+1}\,.
\end{equation}
For $k=3$, we have three transpositions and thus three types of domain walls, each colored differently. The nonzero plaquette terms, also given in \cite{RQCstatmech}, are 
\begin{equation}
\begin{aligned}
&\begin{tikzpicture}[thick,scale=0.5,baseline=0.3cm]
\draw (0,0) -- (1,1.6) -- (-1,1.6) -- (0,0);
\draw[red] (-0.8,0.2) -- (0,1) -- (0,2);
\end{tikzpicture}
= \frac{q}{q^2+1}\,,
&\qquad&
\begin{tikzpicture}[thick,scale=0.5,baseline=0.3cm]
\draw (0,0) -- (1,1.6) -- (-1,1.6) -- (0,0);
\draw[blue] (-0.8,0.2) -- (-0.1,1) -- (-0.1,2);
\draw[red] (0.8,0.2) -- (0.1,1) -- (0.1,2);
\end{tikzpicture}
= \frac{q^2(q^2-1)}{(q^2+2)(q^2+1)(q^2-2)}\,,&\\
&\begin{tikzpicture}[thick,scale=0.5,baseline=0.3cm]
\draw (0,0) -- (1,1.6) -- (-1,1.6) -- (0,0);
\draw[blue] (-0.8,0.2) -- (-0.08,1.03) -- (-0.08,2);
\draw[green] (-0.7,0.1) -- (0.08,1.03) -- (0.08,2);
\end{tikzpicture}
= \frac{q^2(q^2-2)-2}{(q^2+2)(q^2+1)(q^2-2)}\,,
&\qquad&
\begin{tikzpicture}[thick,scale=0.5,baseline=0.3cm]
\draw (0,0) -- (1,1.6) -- (-1,1.6) -- (0,0);
\draw[blue] (-0.8,0.2) -- (-0.08,1.03) -- (-0.08,2);
\draw[red] (0.8,0.2) -- (0.08,1.03) -- (0.08,2);
\draw[green] (-0.7,0.1) -- (0,0.9) -- (0.7,0.1);
\end{tikzpicture}
= -\frac{2(q^2-1)}{(q^2+2)(q^2+1)(q^2-2)}\,,&
\end{aligned}
\end{equation}
as well as reflections and all possible colorings, \eg there are three single domain wall terms for each of the three transpositions

For $k=4$, the generating set for $S_4$ consists of $6$ transpositions. In addition to the same plaquette terms for $k=3$, there are 11 additional plaquette terms for $k=4$ corresponding to different domain wall interactions. For instance, 4 of these terms can be represented as
\begin{align}
&\begin{tikzpicture}[thick,scale=0.5,baseline=0.3cm]
\draw (0,0) -- (1,1.6) -- (-1,1.6) -- (0,0);
\draw[red] (-0.9,0.3) -- (-0.15,1.05) -- (-0.15,2);
\draw[teal] (-0.8,0.2) -- (0,1) -- (0,2);
\draw[blue] (0.8,0.2) -- (0.15,1) -- (0.15,2);
\end{tikzpicture}
=\frac{(q^2-1) (q^4+2 q^2+2)}{(q^2+3)(q^2+2)(q^2+1)(q^2-2)q}\,,\quad&
&\begin{tikzpicture}[thick,scale=0.5,baseline=0.3cm]
\draw (0,0) -- (1,1.6) -- (-1,1.6) -- (0,0);
\draw[teal] (-0.99,0.3) -- (-0.24,1.05) -- (-0.25,2);
\draw[red] (-0.88,0.2) -- (-0.08,1) -- (-0.08,2);
\draw[green] (0.88,0.2) -- (0.08,1) -- (0.08,2);
\draw[violet] (0.99,0.3) -- (0.24,1.05) -- (0.24,2);
\end{tikzpicture}
=\frac{q^2-1}{(q^2+3)(q^2+1)(q^2-3)}\,,& \\
&\begin{tikzpicture}[thick,scale=0.5,baseline=0.3cm]
\draw (0,0) -- (1,1.6) -- (-1,1.6) -- (0,0);
\draw[blue] (-0.9,0.3) -- (-0.15,1.05) -- (-0.15,2);
\draw[red] (-0.8,0.2) -- (0,1) -- (0,2);
\draw[violet] (-0.7,0.1) -- (0.15,0.95) -- (0.15,2);
\end{tikzpicture}
=\frac{q^4-4q^2-1}{(q^2+3)(q^2+1)(q^2-2)q}\,,&
&\begin{tikzpicture}[thick,scale=0.5,baseline=0.3cm]
\draw (0,0) -- (1,1.6) -- (-1,1.6) -- (0,0);
\draw[blue] (-0.8,0.2) -- (-0.08,1.03) -- (-0.08,2);
\draw[red] (0.8,0.2) -- (0.08,1.03) -- (0.08,2);
\draw[green] (-0.7,0.1) -- (0,0.9) -- (0.7,0.1);
\draw[teal] (0.9,0.3) -- (0.24,1.08) -- (0.24,2);
\end{tikzpicture}
= -\frac{(2q^2+1)(q^2-1)}{(q^2+3)(q^2+2)(q^2+1)(q^2-2)q}\,.& \nonumber
\end{align}
Note that the terms with $l$ ingoing domain walls contribute at order $1/q^l$. We also note that none of the plaquette terms have poles at physical values of the local dimension $q\in \mathbb{N}$. This follows from the discussion in App.~\ref{app:designs}, that the poles in the Weingarten function will cancel for physical values of the dimension \cite{Collins04}. We have computed all plaquette terms up to $k=6$ and some additional plaquette terms up to $k=8$, and have verified both the expected asymptotic behavior in $1/q$ and cancellation of poles at $q\in \mathbb{N}$.

\bibliographystyle{utphys}
\bibliography{HaarRQC}

\end{document}